\documentclass[twocolumn]{aastex6}
\usepackage{graphicx}
\usepackage{natbib}
\bibpunct{(}{)}{;}{a}{}{,} 
\usepackage[varg]{txfonts}
\usepackage{hyperref}

\usepackage{color}

\newcommand{\snr} {\mbox{S/N}}

\newcommand{\ie}{i.\,e.}

\newcommand{\eg}{e.\,g.}

\newcommand{\teff}{$T_{{\rm eff}}$}
\newcommand{\kms}{\mbox{km\,s$^{-1}$}}
\newcommand{\ms}{\mbox{m s$^{-1}$}}

\newcommand{\vsini} {$v$\,sin\,$i$}

\newcommand{\vmicro} {\mbox{$\xi_{\rm t}$}}
\newcommand{\gfeh} {\mbox{$[{\rm Fe}/{\rm H}]$}}
\newcommand{\meh} {\mbox{$[{\rm M}/{\rm H}]$}}
\newcommand{\logg} {\mbox{log\,{\it g}}}

\newcommand{\mplanet}{\mbox{$M_{\rm p}$}}
\newcommand{\rplanet}{\mbox{$R_{\rm p}$}}

\newcommand{\mearth}{\mbox{M$_\oplus$}}
\newcommand{\rearth}{\mbox{R$_\oplus$}}
\newcommand{\rhoearth}{\mbox{$\rho_\oplus$}}

\newcommand{\msun}{\mbox{M$_\odot$}}
\newcommand{\rsun}{\mbox{R$_\odot$}}
\newcommand{\mstar}{\mbox{$M_\star$}}
\newcommand{\rstar}{\mbox{$R_\star$}}
\newcommand{\lstar}{\mbox{$\log (L_\star / L_\odot )$}}
\newcommand{\rhostar}{\mbox{$\rho_\star$}}
\newcommand{\rhosun}{\mbox{$\rho_\odot$}}

\newcommand{\logRHK}{\mbox{$\log {\rm R}^{\prime}_{\rm HK}$}}
\newcommand{\SHK}{\mbox{S$_{\rm HK}$}}

\newcommand{\starname}{\mbox{K2-141}}
\newcommand{\planetb}{\mbox{K2-141b}}
\newcommand{\planetc}{\mbox{K2-141c}}
\usepackage[T1]{fontenc}
\usepackage{ae,aecompl}

\begin{document}

\title{An ultra-short period rocky super-Earth with a secondary eclipse and a Neptune-like companion around \starname }

\author{
Luca Malavolta\altaffilmark{1,2},
Andrew W. Mayo\altaffilmark{3,4},
Tom Louden\altaffilmark{5},
Vinesh M. Rajpaul\altaffilmark{6},
Aldo S. Bonomo\altaffilmark{7},
Lars A. Buchhave\altaffilmark{4},
Laura Kreidberg\altaffilmark{3,8},
Martti H. Kristiansen\altaffilmark{9,10},
Mercedes Lopez-Morales\altaffilmark{3},
Annelies Mortier\altaffilmark{11},
Andrew Vanderburg\altaffilmark{3,12,27},
Adrien Coffinet\altaffilmark{13},
David Ehrenreich\altaffilmark{13},
Christophe Lovis\altaffilmark{13},
Francois Bouchy\altaffilmark{13},
David Charbonneau\altaffilmark{3},
David R. Ciardi\altaffilmark{14},
Andrew Collier Cameron\altaffilmark{11},
Rosario Cosentino\altaffilmark{15},
Ian J. M. Crossfield\altaffilmark{16,17,27},
Mario Damasso\altaffilmark{7},
Courtney D. Dressing\altaffilmark{18},
Xavier Dumusque\altaffilmark{13},
Mark E. Everett\altaffilmark{19},
Pedro Figueira\altaffilmark{20},
Aldo F. M. Fiorenzano\altaffilmark{15},
Erica J. Gonzales\altaffilmark{16,28}, 
Rapha\"elle D. Haywood\altaffilmark{3,27},
Avet Harutyunyan\altaffilmark{15},
Lea Hirsch\altaffilmark{18},
Steve B. Howell\altaffilmark{21},
John Asher Johnson\altaffilmark{3},
David W. Latham\altaffilmark{3},
Eric Lopez\altaffilmark{22},
Michel Mayor\altaffilmark{13},
Giusi Micela\altaffilmark{23},
Emilio Molinari\altaffilmark{15,24},
Valerio Nascimbeni\altaffilmark{1,2},
Francesco Pepe\altaffilmark{13},
David F. Phillips\altaffilmark{3},
Giampaolo Piotto\altaffilmark{1,2},
Ken Rice\altaffilmark{25},
Dimitar Sasselov\altaffilmark{3},
Damien S\'egransan\altaffilmark{13},
Alessandro Sozzetti\altaffilmark{7},
St\'ephane Udry\altaffilmark{13},
Chris Watson\altaffilmark{26}
}

\altaffiltext{1}{Dipartimento di Fisica e Astronomia ``Galileo Galilei", Universit\`a di Padova, Vicolo dell'Osservatorio 3, 35122 Padova, Italy; \href{mailto:luca.malavolta@unipd,it}{luca.malavolta@unipd,it}}
\altaffiltext{2}{INAF - Osservatorio Astronomico di Padova, Vicolo dell'Osservatorio 5, 35122 Padova, Italy}
\altaffiltext{3}{Harvard-Smithsonian Center for Astrophysics, 60 Garden Street, Cambridge, MA 02138, USA}
\altaffiltext{4}{Centre for Star and Planet Formation, Niels Bohr Institute \& Natural History Museum, University of Copenhagen, DK-1350 Copenhagen, Denmark}
\altaffiltext{5}{Department of Physics, University of Warwick, Gibbet Hill Road, Coventry CV4 7AL, UK}
\altaffiltext{6}{University of Cambridge, Astrophysics Group, Cavendish Laboratory, J. J. Thomson Avenue, Cambridge CB3 0HE, UK}
\altaffiltext{7}{INAF - Osservatorio Astrofisico di Torino, via Osservatorio 20, 10025 Pino Torinese, Italy}
\altaffiltext{8}{The Harvard Society of Fellows, 78 Mt. Auburn St., Cambridge, MA 02138, USA}
\altaffiltext{9}{DTU Space, National Space Institute, Technical University of Denmark, Elektrovej 327, DK-2800 Lyngby, Denmark}
\altaffiltext{10}{Brorfelde Observatory, Observator Gyldenkernes Vej 7, DK-4340 T\o{}ll\o{}se, Denmark}
\altaffiltext{11}{Centre for Exoplanet Science, SUPA, School of Physics and Astronomy, University of St Andrews, St Andrews KY16 9SS, UK}
\altaffiltext{12}{Department of Astronomy, The University of Texas at Austin, 2515 Speedway, Stop C1400, Austin, TX 78712}
\altaffiltext{13}{Observatoire Astronomique de l'Universit\'e de Gen\`eve, 51 ch. des Maillettes, 1290 Versoix, Switzerland}
\altaffiltext{14}{Caltech/IPAC-NExScI, Mail Code 100-22, 1200 E. California Boulevard, Pasadena, CA 91125, USA}
\altaffiltext{15}{INAF - Fundaci\'on Galileo Galilei, Rambla Jos\'e Ana Fernandez P\'erez 7, 38712 Bre\~na Baja, Spain}
\altaffiltext{16}{Department of Astronomy and Astrophysics, University of California, Santa Cruz, CA 95064, USA}
\altaffiltext{17}{Department of Physics, Massachusetts Institute of Technology, Cambridge, MA 02139, USA}
\altaffiltext{18}{Astronomy Department, University of California Berkeley, Berkeley, CA 94720-3411, USA}
\altaffiltext{19}{National Optical Astronomy Observatory, 950 N. Cherry Avenue, Tucson, AZ 85719, USA}
\altaffiltext{20}{Instituto de Astrof\' isica e Ci\^encias do Espa\c{c}o, Universidade do Porto, CAUP, Rua das Estrelas, PT4150-762 Porto, Portugal}
\altaffiltext{21}{NASA Ames Research Center, Moffett Field, CA 94035, USA}
\altaffiltext{22}{NASA Goddard Space Flight Center, 8800 Greenbelt Rd, Greenbelt, MD 20771, USA}
\altaffiltext{23}{INAF - Osservatorio Astronomico di Palermo, Piazza del Parlamento 1, 90124 Palermo, Italy}
\altaffiltext{24}{INAF - Osservatorio Astronomico di Cagliari, Via della Scienza 5 - 09047 Selargius (CA)}
\altaffiltext{25}{SUPA, Institute for Astronomy, University of Edinburgh, Royal Observatory, Blackford Hill, Edinburgh, EH93HJ, UK}
\altaffiltext{26}{Astrophysics Research Centre, School of Mathematics and Physics, Queen's University Belfast, Belfast BT7 1NN, UK}
\altaffiltext{27}{NASA Sagan Fellow}
\altaffiltext{28}{National Science Foundation Graduate Research Fellow}

\begin{abstract}
Ultra-short period (USP) planets are a class of low mass planets with periods shorter than one day. Their origin is still unknown, with photo-evaporation of mini-Neptunes and in-situ formation being the most credited hypotheses. Formation scenarios differ radically in the predicted composition of USP planets, it is therefore extremely important to increase the still limited sample of USP planets with precise and accurate mass and density measurements. We report here the characterization of an USP planet with a period of $0.28$ days around \starname\ (EPIC~246393474), and the validation of an outer planet with a period of $7.7$ days in a grazing transit configuration. We derived the radii of the planets from the {\it K2} light curve and used high-precision radial velocities gathered with the HARPS-N spectrograph for mass measurements. For \planetb\ we thus inferred a radius of $1.51\pm0.05$~\rearth\ and a mass of $5.08\pm0.41$~\mearth, consistent with a rocky composition and lack of a thick atmosphere. \planetc\ is likely a Neptune-like planet, although due to the grazing transits and the non-detection in the RV dataset, we were not able to put a strong constraint on its density. We also report the detection of secondary eclipses and phase curve variations for \planetb. The phase variation can be modelled either by a planet with a geometric albedo of $0.30 \pm 0.06$ in the {\it Kepler} bandpass, or by thermal emission from the surface of the planet at $\sim$3000K. Only follow-up observations at longer wavelengths will allow us to distinguish between these two scenarios.
\end{abstract}

\shorttitle{The \starname\ system}
\shortauthors{Malavolta et al.}

\maketitle

\section{Introduction}\label{sec:introduction}
The origin of ultra-short period (USP) planets, \ie , planets with periods shorter than one day and radii smaller than 2 \rearth , is still unclear. An early hypothesis suggested that USP planets and small planets in general were originally Hot Jupiters (HJs) that underwent strong photo-evaporation due to the high insolation flux, \citep[\eg , thousands of times that of Earth,][]{LecavelierdesEtangs2004} ending up with the complete removal of their gaseous envelope and their solid core exposed. The paucity of gas giants observed in the photo-evaporation desert, \ie , the region around a star where only solid cores of once-gaseous planets could survive, is the most convincing proof of evaporation as a viable process to form small planets \cite[\eg ,][]{LecavelierdesEtangs2007,Davis2009,Ehrenreich2011,Beauge2013}. In the case of USP planets, \cite{SanchisOjeda2014} also found an occurrence rate of USP planets similar to that of HJs using data from the {\it Kepler} mission, but recently, thanks to Keck spectroscopy on a magnitude-limited subset of the same sample, \cite{Winn2017} discovered that the metallicity distributions of the two populations are significantly different, thus rejecting the idea of a common origin. The same study supports a similar hypothesis in which the progenitors of USP planets are not the HJs but the so-called mini-Neptunes, \ie , planets with rocky cores and hydrogen-helium envelopes, typically with radii between 1.7 and 3.9 \rearth\ and masses lower than $\sim  10$ \mearth .
An origin of USP planets as photo-evaporated mini-Neptunes is also consistent with the lack of planets with radii between 2.2 and 3.8 \rearth\ with incident flux higher than 650 times the Solar constant \citep{Lundkvist2016}, the gap between 1.5 and 2 \rearth\ in the population of planets with periods shorter than 100 days \citep{Fulton2017}, and the multiplicity of USP planets, typically found with small companions at longer periods \citep{SanchisOjeda2014}.
While observations of known HJs have confirmed the stability of their atmospheres against evaporation \citep[starting from ][]{Vidal-Madjar2003}, and theory has always struggled to explain the strong photo-evaporation that HJs should undergo to become USP planets \citep[\eg ][]{MurrayClay2009}, removing the outer envelope of a mini-Neptune is theoretically less challenging and several models have successfully reproduced the properties of observed USP planets using either photo-evaporation \citep[\eg ,][]{Lopez2017} or improved models for Roche lobe overflow \citep[\eg ,][]{Jackson2017}, in agreement with observations of mini-Neptunes undergoing evaporation \citep{Ehrenreich2015}. Alternatively, USP planets may represent the short-period tail of the distribution of close-in rocky planets migrated inwards from more distant orbits \citep[\eg ,][]{LeeChiang2017} or formed in-situ \citep[\eg ,][]{Chiang2013}, although the latter hypothesis would have difficulties explaining the presence of thick envelopes accreted within the snow line.

It appears clear that only a systematic study of the internal and atmospheric composition of USP planets, in conjunction with the amount of irradiation to which they are subjected and the presence of other companions in the system, can shed light on their origin. In order to do so, we need precise and accurate measurements of both their radius and mass. Most of the {\it Kepler} and {\it K2} USP candidates orbit stars too faint for precise radial velocity (RV) follow-up, and so far only a handful of USP planets have reliable density estimates.

In addition to discovering most of the USP planets known to date, the excellent quality of {\it Kepler} data has also revealed the secondary eclipse and phase variations of two of them, namely Kepler-10b \citep{Batalha2011} and Kepler-78b \citep{SanchisOjeda2013}. If USP planets were really lava-ocean worlds, their atmospheres would be likely made of heavy-element vapors with a very low pressure and, being tidally locked, would experience extremely high day-night contrasts \citep{Leger2011}. Consequently, the bottom of the secondary eclipse is expected to be about at the same level as just before/after the primary transit, when only the nightside of the planet is in view. This seems to be the case with Kepler-78b \citep{SanchisOjeda2013} and Kepler-10b \citep{Esteves2015}, even though a non-negligible night-side temperature for the latter has been reported by \cite{Fogtmann-Schulz2014}. The geometric albedos of both planets could not be well constrained because of the degeneracy between thermal and reflected light in the Kepler bandpass, which could be broken with observations of the occultation and phase curve at IR wavelengths \citep[\eg ,][]{Schwartz2015}. Noteworthy is the attempt by \cite{Rouan2011} to use a lava-ocean model to interpret the optical occultation and phase curve of Kepler-10b.

In this paper, we report on the discovery, characterization, and confirmation of an USP planet, and the discovery and validation of an outer companion planet with grazing transits around an active K4  dwarf, \starname\ (EPIC~246393474), discovered in the Campaign 12 data of the {\it K2} mission and then observed with the high-precision HARPS-N spectrograph for radial velocity confirmation. We tackled the determination of mass and radius of the star, which ultimately can affect the planets' properties, using three independent methods for the atmospheric parameters and including any additional data available from the literature. After validating the planets, we measured their masses using three methods that rely on different assumptions for the stellar activity modeling, to ensure that our mass estimates are not biased by a specific choice of stellar activity treatment. We compare the density obtained for \planetb\ with the distribution of USP planets in the mass-radius (M-R) diagram. We also detected the secondary eclipse and phase variations of planet b in the {\it K2} light curve, and used this information to constrain the geometric albedo of the planet and its thermal emission. \footnote{A paper on the validation and mass measurement of \planetb\ has been submitted to {A\&A} by The {\it KESPRINT} consortium while this paper was already in an advanced state of preparation.}

\section{Observations}\label{sec:observations}

\subsection{{\it K2} photometry}\label{sec:kepler-photometry}

\begin{figure}
\includegraphics[width=\linewidth]{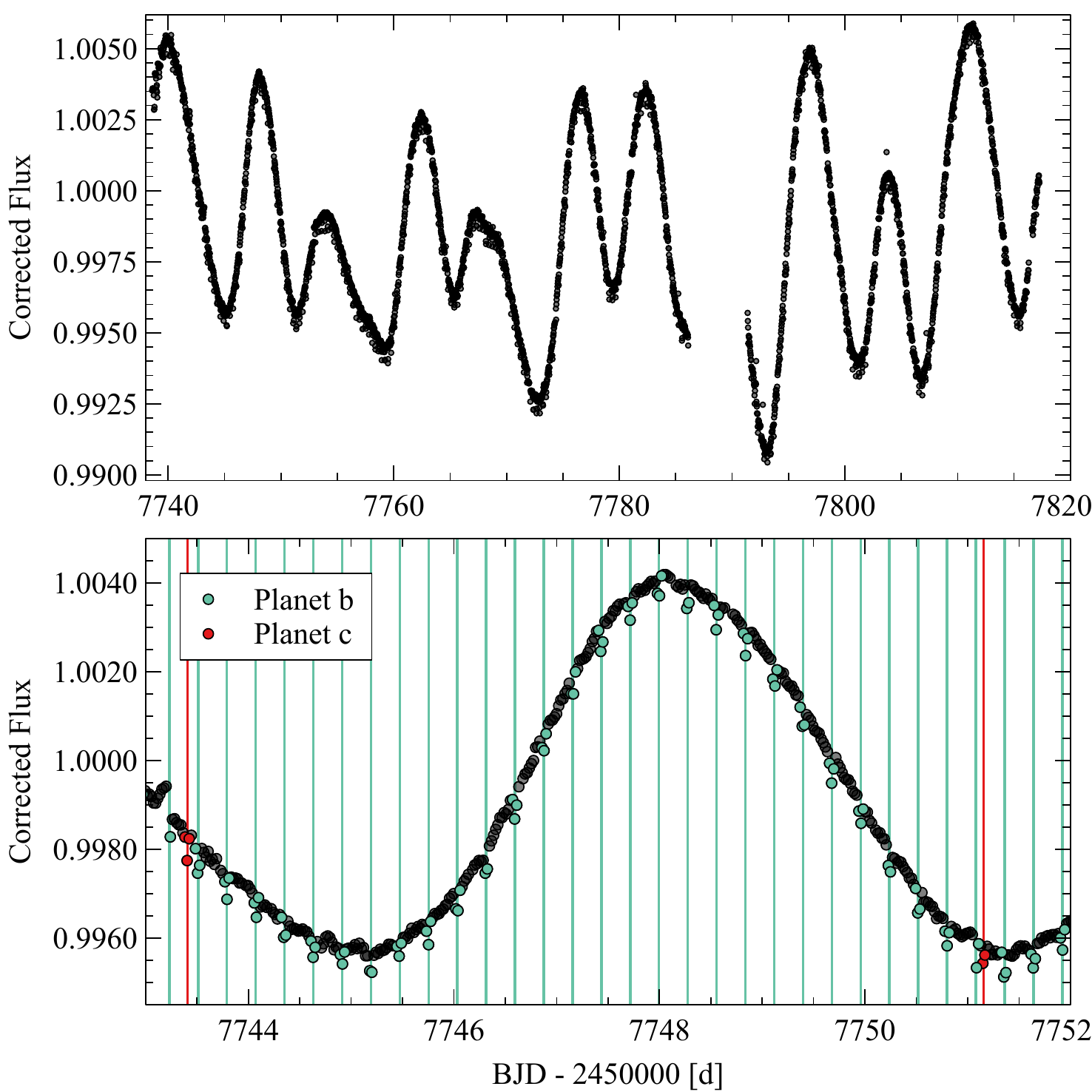}
\caption{{\it Top}: {\it K2} light curve of \starname . {\it Bottom:} A portion of the light curve is shown to highlight the two transiting planets.}
\label{fig:EP246393474_K2_LC}
\end{figure}

\starname\ first came to our attention after it was observed with the {\it Kepler} space telescope during Campaign 12 of its extended {\it K2} mission\footnote{The star was proposed as a target from the following  {\it K2} General Observer programs: 12071, D. Charbonneau; 12049, E. Quintana; 12122, A. Howard; 12123, D. Stello; 12904, {\it K2} GO Office.}. \starname\  was observed by {\it K2} for about 80 days between 15 December 2016 and 4 March 2017, with a loss of 5.3 days of data due to a safe mode state, presumably caused by a reset of flight software. Afterwards, the data were downlinked to Earth, processed by the {\it Kepler} pipeline to calibrate the raw pixel level data, and released publicly. We downloaded the data for \starname\ and all other targets observed by {\it K2} during Campaign 12 from the Mikulski Archive for Space Telescopes (MAST)\footnote{\url{https://archive.stsci.edu/k2/}}, produced light curves from the calibrated pixel files following \cite{Vanderburg2014}, and searched for transits as described by \citet{Vanderburg2016}. Our transit search identified a strong signal at a period of only 6.7 hours. Using {\tt LcTools}\footnote{Available at \url{https://sites.google.com/a/lctools.net/lctools/home}} \citep{Kipping2015}, we color coded this signal in order to enhance the visibility of hidden candidate signals, and a subsequent visual inspection of the {\it K2} light curve revealed a second planet candidate with a period of 7.75 days. The duration of the second transit signal is short and V-shaped -- consistent with a planet transiting in a grazing architecture -- which is likely why our automated search pipeline failed to identify the signal.  We pinpoint a total of nine transits of \planetc\ during the {\it K2} baseline, one transit was lost while {\it Kepler} was in safe mode. All overlaps of the two planets consist of single long cadence data points and none of these are located at mid-transit. We confirmed the periodicity with a subsequent, more thorough analysis following the prescriptions of \cite{Bonomo2012}.
The full {\it K2} light curve is shown in Figure~\ref{fig:EP246393474_K2_LC}. In addition to the two transiting signals, there is a clear modulation (total excursion of 0.015 mmag) most likely due to the stellar activity of the star.

After removing the stellar activity signal from the {\it K2} light curve and phase-folding the data to the orbital period of \planetb, we also identify the signal of the secondary eclipse of this planet, centered around phase 0.5 and with a duration consistent with that of the primary transit. In addition to the eclipse signal, we observe what appears to be modulation of the light curve with phase. We further explore these features in Section~\ref{sec:secondary_analysis}. We repeated this analysis for planet c, and did not find any evidence of a detectable phase curve or secondary eclipse.

\subsection{Radial Velocities}\label{sec:radial_velocities}

We collected 44 spectra using HARPS-N at the Telescopio Nazionale Galileo (TNG), in La Palma \citep{Cosentino2012}, with the goal of precisely determining the mass of the USP planet. To reach this goal we followed a twofold strategy: we gathered at least two points each night (when weather allowed) in order to remove activity variations by applying nightly offsets \citep[\eg ,][]{Hatzes2011,Pepe2013}, and we observed the target for a duration of a few stellar rotations to be able to use Gaussian process  regression \citep[\eg ,][]{Haywood2014, Rajpaul2015} to model the stellar activity signals directly.

At the magnitude of our target ($V=11.5$), HARPS-N delivers an average RV internal error of 2.9 \ms\ for a single exposure of 1800 seconds (average \snr~of~$42 $ at 5500 \AA  ), to be compared with an instrumental stability better than 1 \ms\ \citep{Cosentino2014}. In other words, our error budget is largely dominated by photon noise.

Therefore we chose the {\it objAB} observational setup, \ie , the second fiber (fiber B) observed the sky instead of acquiring a simultaneous Fabry-Perot calibration spectrum to correct for the instrumental RV drift.

Data were reduced using the standard Data Reduction Software (DRS) using a {\tt K5} flux template (the closest match to the spectral type of the target) to correct for variations in the flux distribution as a function of the wavelength, and a {\tt K5} binary mask to compute the cross-correlation function (CCF) \citep{Baranne1996, Pepe2002}. We corrected the spectra for Moon contamination as explained in \cite{Malavolta2017a}, {and found that only two spectra were strongly affected by sky background.
The resulting RV data with their formal 1$\sigma$ uncertainties and the associated activity indices (see Section~\ref{sec:stellar_activity} for more details) are listed in Table~\ref{table:RV_table}.

\begin{table*}
  \caption{HARPS-N Radial Velocity Measurements }
\label{table:RV_table}      
\centering                                      
\begin{tabular}{c c c c c c c c c }          
\tableline\tableline                        
  BJD$_{\rm TDB}$ & RV & $\sigma _{\text RV}$ & BIS & FWHM &  \SHK & $\sigma_{\text \SHK }$ & H$_{\alpha}$ & $\sigma_{\text{H} \alpha }$ \\
 $[$d$]$  & [\ms ] & [\ms ] & [\ms ] & [\kms ] & [dex] & [dex] & [dex] & [dex] \\
\tableline                                 
2457972.6416  &  -3379.6  &  2.3  &  44.1  &  6.955  &  0.964  &  0.019  &  0.2938  &  0.0010 \\
2457989.5731  &  -3383.9  &  3.9  &  51.0  &  6.951  &  0.951  &  0.037  &  0.2901  &  0.0018 \\
2457991.5524  &  -3373.2  &  4.7  &  33.5  &  6.918  &  0.959  &  0.048  &  0.2876  &  0.0012 \\
... & ... & ... & ... & ... & ... & ... & ... & ...\\

\tableline
\end{tabular}
\tablecomments{Table \ref{table:RV_table} is published in its entirety in the machine-readable format. A portion is shown here for guidance regarding its form and content.}
\end{table*}

\section{Stellar parameters}\label{sec:stellar_parameters}

For late-type stars like our target, systematic errors in the stellar photospheric parameters due to different assumptions and theoretical models largely dominate the internal error estimates for the most diffused methods, \eg , see the spread in temperature and metallicity in the case of the bright star HD219134 \citep{Motalebi2015}.
In this work we obtained the stellar photospheric parameters with three complementary methods, and  we assumed $\sigma_{T_{{\rm eff}}} = 100$~K , $\sigma_{\text{log\,{\it g}}} = 0.2$, $\sigma_{[{\rm Fe}/{\rm H}]} = 0.06$ as a good estimate of the systematic errors regardless of the internal error estimates, for all methods. This choice also avoided privileging one technique over the others when deriving the mass and radius of the star.

\paragraph{Empirical calibration} {\tt CCFpams}\footnote{Available at \url{https://github.com/LucaMalavolta/CCFpams}} is a  method based on the empirical calibration of temperature, metallicity and gravity on the equivalent width of CCFs obtained with selected subsets of stellar lines, according to their sensitivity to temperature. We refer the reader to \cite{Malavolta2017b} for more details on this method. CCFs were computed on the individual spectra and then co-added for their equivalent width measurement. We obtained \teff~$ = 4713 $~K, \logg~$ = 4.76$ (after applying the correction from \citealt{Mortier2014}) and \gfeh~$ = -0.15$.

\paragraph{Equivalent widths} The classical curve-of-growth approach consists in deriving temperature and microturbulent velocity \vmicro\ by minimizing the trend of iron abundances (obtained from the equivalent width of each line) with respect to excitation potential and reduced equivalent width respectively, while the gravity \logg\ is obtained by imposing the same average abundance from neutral and ionized iron lines. Equivalent width measurements were carried out with {\tt  ARESv2}\footnote{Available at \url{http://www.astro.up.pt/~sousasag/ares/}} \citep{Sousa2015}, while line analysis and spectrum synthesis was performed using {\tt  MOOG}\footnote{Available at \url{http://www.as.utexas.edu/~chris/moog.html}} \citep{Sneden1973} jointly with the {\tt ATLAS9} grid of stellar model atmosphere from \cite{Castelli2004}, under the assumption of local thermodynamic equilibrium (LTE). We followed the prescription of \cite{Andreasen2017} and applied the gravity correction from \cite{Mortier2014}. The analysis was performed on the resulting coaddition of individual spectra. We obtained  \teff~$ = 4518 $~K, \logg~$ = 4.76$, \gfeh~$ = 0.00$ and \vmicro~$= 0.63 \pm 0.35$~\kms .

\paragraph{Spectral synthesis match} The Stellar Parameters Classification tool ({\tt SPC}, \citealt{Buchhave2012, Buchhave2014}) performs a cross-correlation of the observed spectra with a library of synthetic spectra and then interpolates the resulting correlation peaks to determine the best-matching effective temperature, surface gravity, metallicity and line broadening.
The quoted results are the average of the values measured from each exposure. We obtained \teff~$ = 4622 $~K, \logg~$ = 4.63$, \meh~$= 0.00$ and  \vsini~$ = 1.5 \pm 0.4 $~\kms .

\medskip
We determined the stellar mass and radius using {\tt isochrones} \citep{Morton2015}, with posterior sampling performed by {\tt MultiNest} \citep{Feroz2008,Feroz2009,Feroz2013}. We provided as input the parallax of the target from the Tycho-GAIA Astrometric Solution ($p=17.0 \pm 0.8 $~mas, $d= 59 \pm 3 $~pc, \citealt{GAIAcoll2016a, GAIAcoll2016b}) plus the photometry from the Two Micron All Sky Survey (2MASS, \citealt{Cutri2003,Skrutskie2006}) and the Wide-field Infrared Survey Explorer (WISE, \citealt{Wright2010}). We did not use the GAIA magnitude because it was not consistent with the measured parallax and the wide-band photometry. For stellar models we used both MESA Isochrones \& Stellar Tracks (MIST, \citealt{Dotter2016,Choi2016,Paxton2011}) and the Dartmouth Stellar Evolution Database \citep{Dotter2008}. To assess the influence of the broadband photometry and the different photospheric parameters, for each set of spectroscopic parameters we performed the analysis including both or only one of the photometric sets, for a total of nine posteriors sampling distribution for each parameter.
From the median and standard deviation of all the posterior samplings we obtained  \mstar~$=0.708 \pm 0.028$~\msun\ and $ \rstar = 0.681 \pm 0.018\ \rsun $. We derived the stellar density \rhostar~$ = 2.244 \pm 0.161$~\rhosun\  directly from the posterior distributions of \mstar\ and \rstar\ .
The astrophysical parameters of the star are summarized in Table~\ref{table:stellar_parameters}, where the temperature, gravity and metallicity are those obtained from the posteriors distributions, in a similar fashion to mass and radius, and take into account the constraint from GAIA parallax. The \logRHK\ quoted in the table was obtained using the calibration from \cite{Lovis2011} and assumed $B-V=1.19$ instead of $B-V=0.69$ as listed in the Simbad catalogue \citep{Wenger2000}, which is not consistent with the spectral type of the star. The chosen value is set by the upper limit in the calibration, which is however well within the error bars of the outcome of the isochrone fit, $B-V = 1.21 \pm 0.20$. Nevertheless, this estimate of \logRHK\ should be taken with caution.

\begin{table}
  \caption{Astrophysical parameters of the star.}
\label{table:stellar_parameters}      
\centering                                      
\begin{tabular}{l c c }          
\tableline\tableline                        
  Parameter & Value & Unit  \\
  \tableline                                 
  \noalign{\smallskip}
  EPIC number & 246393474 & \\
  2MASS alias & J23233996-0111215 & \\
  $\alpha_{\rm J2000}$ & 23:23:39.97 & hms \\
  $\delta_{\rm J2000}$ & -01:11:21.39 & dms \\
  \rstar\  & $ 0.681 \pm 0.018 $ & \rsun \\
  \mstar\ & $ 0.708 \pm 0.028 $ & \msun \\
  \rhostar & $ 2.244 \pm 0.161$ & \rhosun  \\
  \lstar\  & $-0.75 \pm 0.04 $ & - \\
  \teff\ & $ 4599 \pm 79 $ & K \\
  \logg\  & $ 4.62_{-0.03}^{+0.02}$ &-  \\
  \gfeh\  & $ -0.06_{-0.10}^{+0.08}$ & - \\
  distance & $ 61 \pm 2 $ & pc \\
  A$_V$  & $0.14_{-0.10}^{+0.14}$ & mag \\
  age  & $ 6.3_{-4.7}^{+6.6}$ & Gy \\
  \logRHK & $-4.6 \pm 0.1$ & - \\
\tableline
\end{tabular}
\end{table}

\section{Stellar activity}\label{sec:stellar_activity}
The precise and continuous coverage provided by {\it K2} photometry offers the best chance to determine the stellar rotation period and put a lower limit to the decay time scale of the active regions. In the following, we performed the analyis on the {\it K2} light curve after removing those points affected by a transit, using the solution in Section~\ref{sec:photometric_analysis}.

The Generalized Lomb-Scargle (GLS, \citealt{Zechmeister2009}) periodogram of the light curve and the bisector inverse span (BIS) detected a main periodicity around 7 days. The spectroscopic activity diagnostics, namely the Full Width Half Maximum (FWHM) of the CCF, the \logRHK\ index \citep{Lovis2011}, and the  H$\alpha$ index \citep{GomesDaSilva2011,Robertson2013}, however, did not confirm this result, all suggesting instead a main periodicity around 14 days (Figure~\ref{fig:GLS_periodograms_all}). The auto correlation function on the {\it K2} data, computed as described in \cite{McQuillan2013}\footnote{As implemented in \url{https://github.com/bmorris3/interp-acf}} also converged to 14 days.
We note that the lack of precise photometry in the $B$ and $V$ bands prevented us from determining an accurate \logRHK\, so we decided to analyze the \SHK\ index instead.
\begin{figure}
\includegraphics[width=\linewidth]{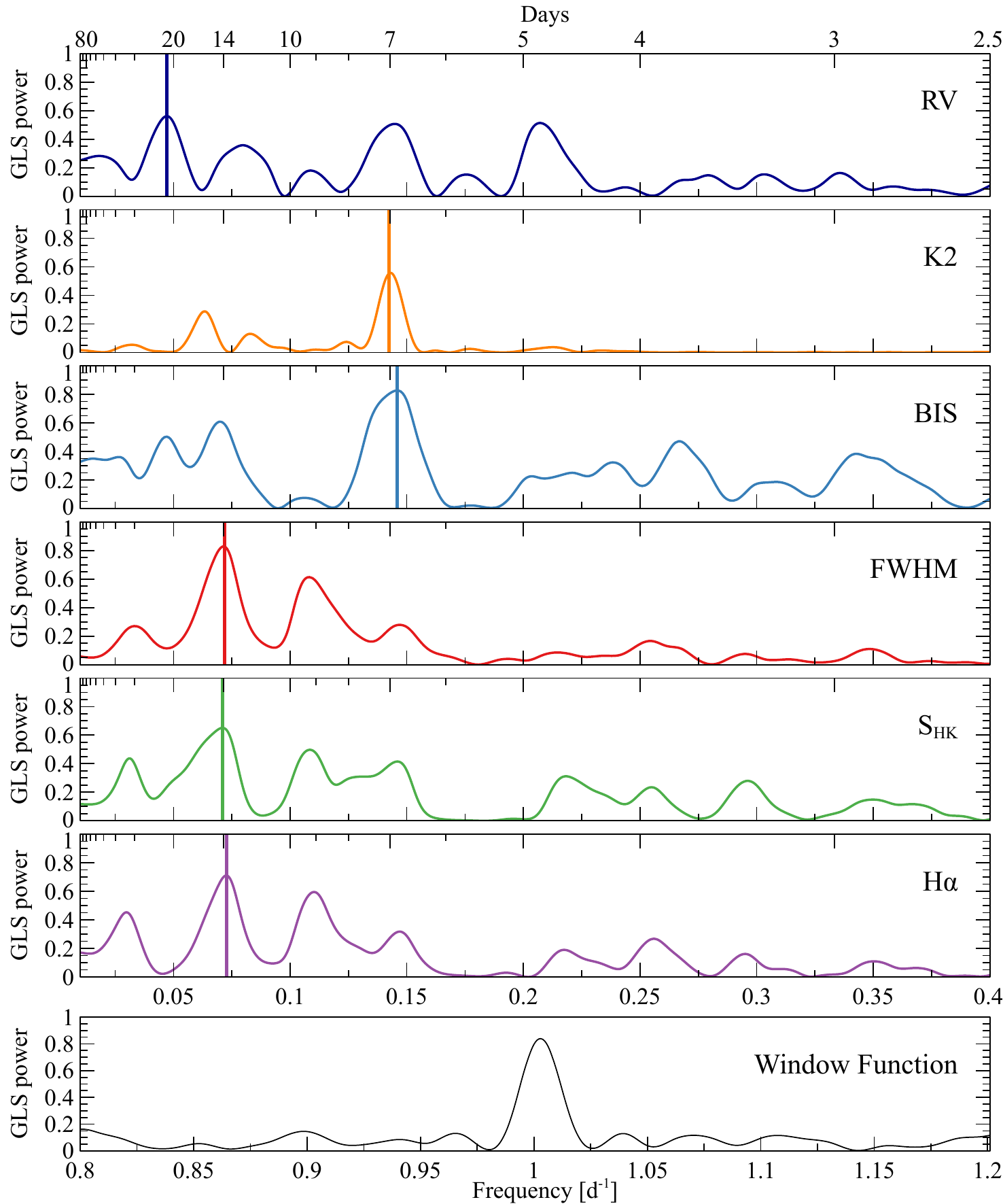}
\caption{Generalized Lomb-Scargle periodogram of the RVs, the {\it K2} light curve and spectroscopic activity indices. A first analysis of the {\it K2} data and bisector inverse span (BIS) returns a main periodicity around 7 days, which could be mistaken as the rotational period of the star if the other activity indices are not considered. Subsequent analysis confirmed a rotational period around 14 days.}
\label{fig:GLS_periodograms_all}
\end{figure}

An accurate value for the rotational period of the star is of paramount importance for the correction of activity induced signals. To understand the disagreement between the {\it K2} light curves and the activity indices we followed the recipe of \cite{2017arXiv170605459A}, who suggest a Gaussian process (GP) with a  quasi-period covariance kernel function as a more reliable method than those mentioned above to measure rotational periods of active stars. We performed our analysis using version 5 of {\tt PyORBIT\footnote{Available at \url{https://github.com/LucaMalavolta/PyORBIT}}} \citep{Malavolta2016}, a package for RV and activity indices analysis, with the implementation of the GP quasi-period kernel as described in \cite{Grunblatt2015}, from which we inherit the mathematical notation, through the {\tt george\footnote{Available at \url{https://github.com/dfm/george}}} package \citep{Ambikasaran2015}.

Since GP regression ordinarily scales with the third power of the number of data points, to ease the analysis of the {\it K2} dataset we binned the light curve every 3--4 points, paying attention that all the points within a bin belonged to the same section within two transits and checking that the binning process did not alter the overall shape of the light curve. For the activity indices this step was not required. We obtained
$P_{\rm rot} = 13.9 \pm 0.2 $ d from the {\it K2} light curve,
$P_{\rm rot} = 13.7 \pm 0.2 $ d from the BIS, and similar values from all the other activity indices,
thus confirming that the peak seen in the GLS periodograms of {\it K2} and BIS corresponds to the first harmonic of the true rotational period. The decay time scale of active regions  $\lambda$ and the coherence scale $w$ were constrained only in the {\it K2} data, with  $\lambda = 12.8 \pm 1.0 $ d and $ w = 0.34 \pm 0.02 $, and a covariance amplitude of $h_{\rm K2} = 0.0031 \pm 0.0004$~mag.
Finally, we note that despite the high level of activity of the star, no evident correlation is seen between the RV and the activity indices (Figure~\ref{fig:GLS_correlations}), meaning that a simple linear correlation model would likely fail in removing the activity signal from the RV observations.

\begin{figure}
\includegraphics[width=\linewidth]{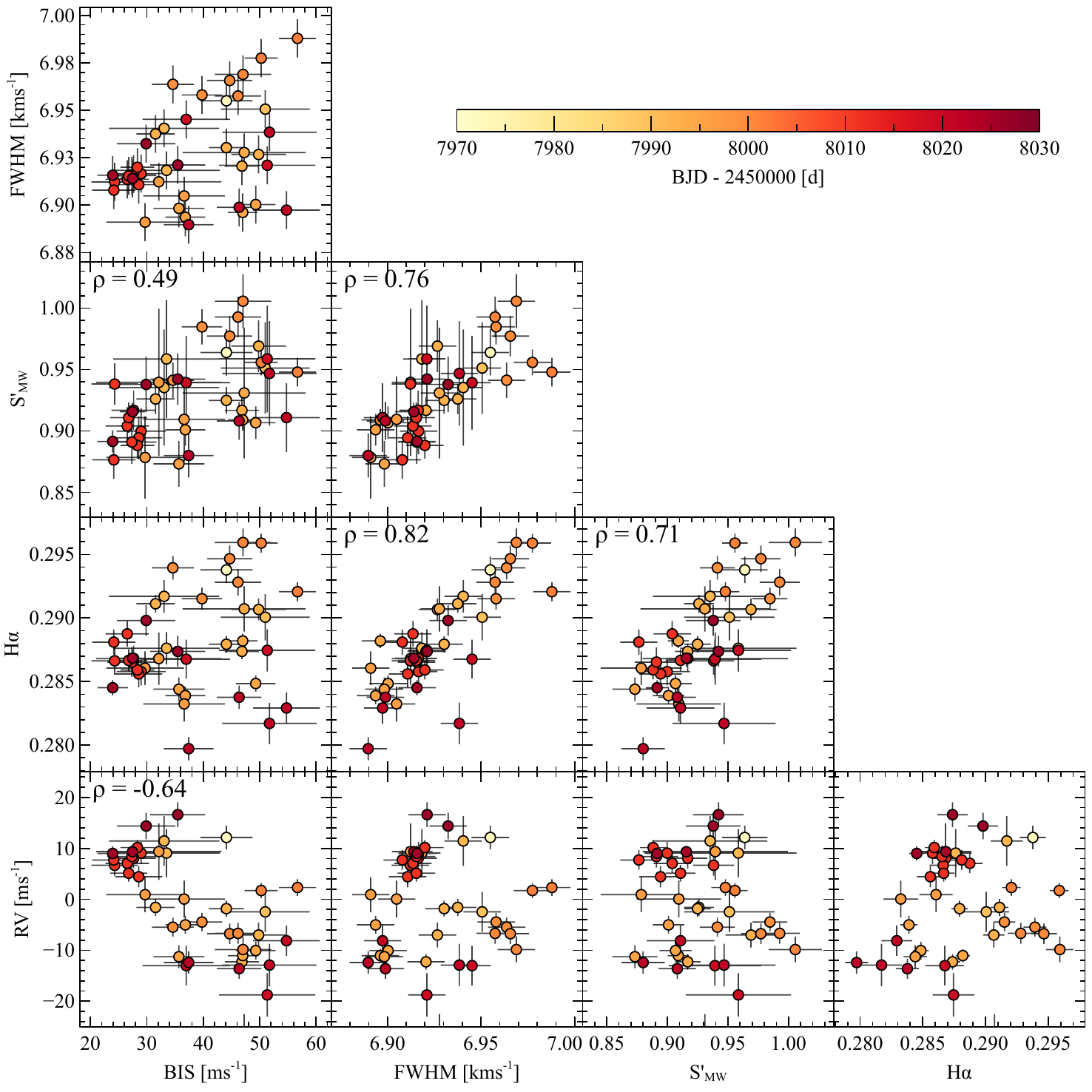}
\caption{Corner plot of the activity indices and RVs. The contribution of planet b has been removed from the RVs to highlight the correlation with the activity indices. The Pearson correlation coefficient $\rho$ is reported only when its $p$-value is lower than $10^{-2}$. FWHM, \SHK\ and H$\alpha$ are strongly correlated with each other but only weakly with the RVs, suggesting that a more complex model to correct for stellar activity is required.}
\label{fig:GLS_correlations}
\end{figure}

\section{Planets validation}\label{sec:planets-validation}
Both planets were subjected to a validation procedure in order to calculate the false positive probability (FPP) for each planet. The full details of the analysis will be described in Mayo~et~al.~(submitted). Here we give a brief summary for the reader's convenience. Our validation process was conducted with Validation of Exoplanet Signals using a Probabilistic Algorithm, or {\tt vespa}. {\tt vespa} is a public package \citep{Morton2015b} based on the work of \citet{Morton2012}. It analyzes input information such as sky position, parallax, stellar parameters, broadband photometry, light curve shape, and contrast curves. {\tt vespa} then creates a representative stellar sample for the true positive scenario and each false positive scenario (i.e. eclipsing binaries, background eclipsing binaries, and hierarchical eclipsing binaries). For each scenario, the sample is cut down to the subset of systems which reproduce the input observations. Finally, the ratio between the number of remaining false positive scenarios and the number of total remaining scenarios is returned as the FPP. In our case, for each planet we provided {\tt vespa} with the equatorial coordinates, a GAIA parallax, stellar photospheric parameters (\teff , \logg , \gfeh ), J, H, and K broadband photometry from 2MASS, a normalized light curve (with the other planet's transits removed), and three contrast curves extracted from one adaptive optics image and two simultaneous speckle images collected at the 3-m Lick Observatory telescope and using NESSI at the 3.5-m WIYN Observatory telescope respectively \citep[][Scott et al. in prep.]{Howell2011}.
The speckle and adaptive optics images were obtained from the Exoplanet Follow-up Observing Program (ExoFOP) for {\it K2} website\footnote{\url{https://exofop.ipac.caltech.edu/k2/}}.

After {\tt vespa} calculated the probability of different scenarios, we applied an additional constraint based on our RV observations with HARPS-N. Our numerous HARPS-N observations conclusively ruled out scenarios where the transit signals we see are caused by an eclipsing binary on the foreground star. We therefore reduced the probability of these scenarios to 0. We also took into account the fact that there are multiple transit signals detected in the direction of \starname. Statistically, candidates around stars hosting more than one possible transit signal are considerably more likely to be genuine exoplanets than those in single-candidate systems\citep{Latham2011, Lissauer2012}. To take this into account, we divided both FPPs by 25 (\citealt{Lissauer2012}, see also \citealt{Sinukoff2016} and \citealt{Vanderburg2016c} who estimated this factor for K2 candidates.). After including these constraints and factors, we calculated false positive probabilities of FPP $< 10^{-4}$ for planet b, and FPP $= 4.8\times10^{-4}$ for planet c. These false positive probabilities are low enough that we consider planet c to be statistically validated, while we consider planet b to be {\em confirmed} by our detection of its spectroscopic orbit with HARPS-N.

\section{Photometric Analysis}\label{sec:photometric_analysis}

After we had identified the two candidate signals, we reprocessed the {\it K2} light curve by simultaneously fitting for the {\it K2} flat field systematics, transit light curves, and stellar variability using the procedure described by \citet{Vanderburg2016}. The final light curve is shown in Figure~\ref{fig:transit_plot}.

\begin{figure}
\includegraphics[width=\linewidth]{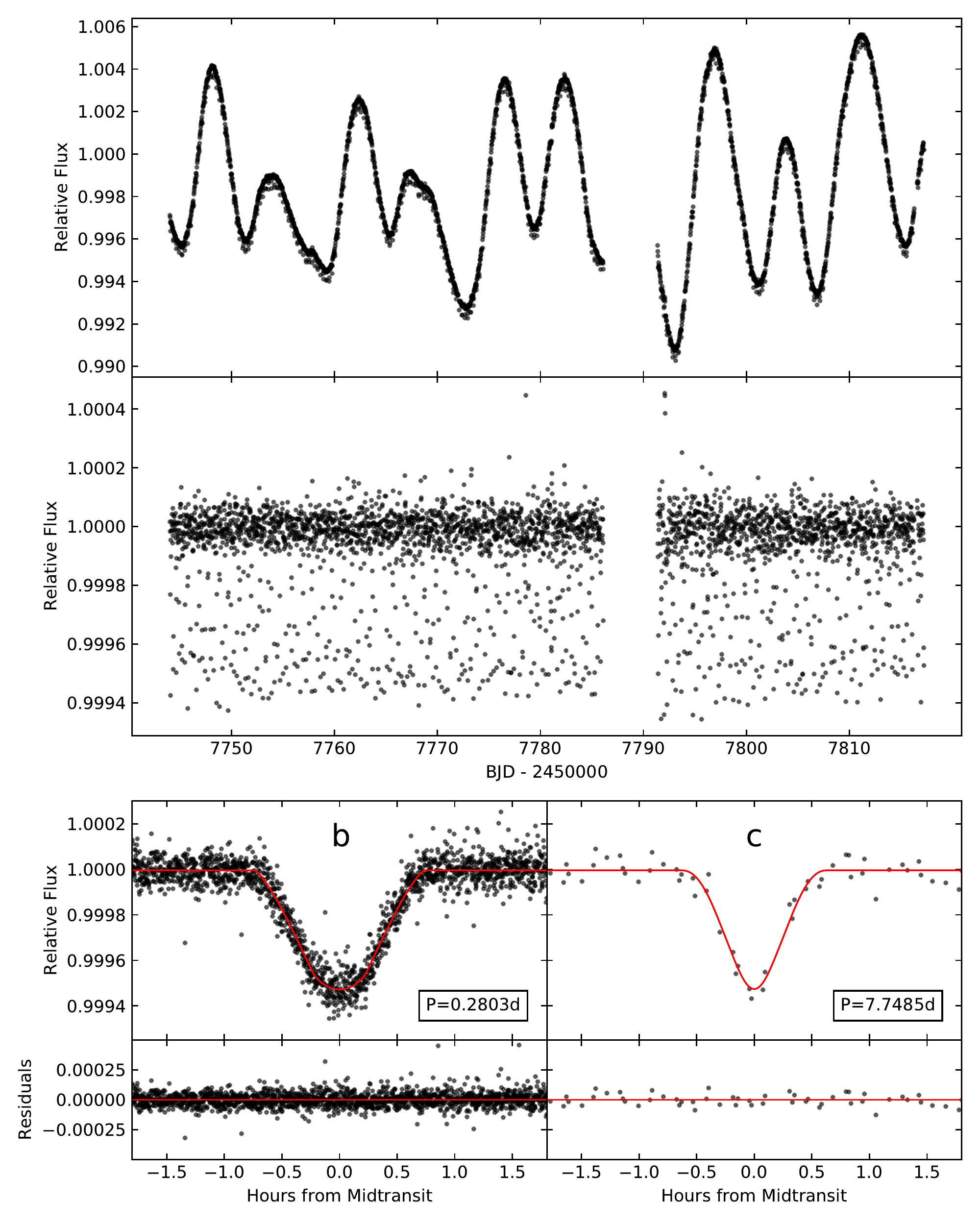}
\caption{{\it Top:} Systematics-corrected and normalized {\it K2} light curve (top and bottom panel respectively). {\it Bottom:} The phase-folded light curve for planets b and c with model in red (residuals in panels below)}
\label{fig:transit_plot}
\end{figure}

We modeled the normalized light curve using the {\tt batman} transit model \citep{Kreidberg2015}. We assumed the planets were non-interacting with zero eccentricity orbits. We also accounted for the long cadence integration by including an exposure time of 1764.944s in the model \citep{Kipping2010,Swift2015}. The model included a baseline flux offset parameter, a noise parameter (since the \citealt{Vanderburg2014} reduction method does not produce flux uncertainties), and two quadratic limb-darkening parameters \citep{Kipping2013}. Further, each of the two planets was modeled with five parameters: the epoch (i.e. time of first transit), period, inclination, planetary to stellar radius ratio ($R_\mathrm{p}/$\rstar ), and semi-major axis normalized to the stellar radius ($a/$\rstar ). Parameters and their uncertainties were estimated using a Markov chain Monte Carlo (MCMC) algorithm with an affine invariant ensemble sampler \citep{GoodmanWeare2010}. We implemented the simulation via the {\tt emcee} Python package \citep{ForemanMackey2013} and ran it with a 28 chain ensemble (twice the number of model parameters). Our model parameters and uncertainties were estimated upon convergence, which we defined as the point in the MCMC simulation when the scale-reduction factor \citep{GelmanRubin1992} was $< 1.1$ for all parameters. The simulation assumed a uniform prior for all parameters except $R_\mathrm{p}/$\rstar , for which we applied a log-uniform prior. We also calculated stellar density at each simulation step (by using period and $a/$\rstar\ to solve for density with Kepler's third law) and applied a prior penalty by comparing it to our estimate of spectroscopic density and its uncertainties ($2.244 \pm 0.161$ \rhosun ).

\begin{table}
\caption{Planet parameters from {\it K2} light curve and RV fitting}
\label{table:K2_parameter_planets}      
\begin{tabular}{l c c }          
  \tableline\tableline                        
  Parameter & \planetb & \planetc \\
  \tableline
  $P$ [d] &  $0.2803244 \pm  0.0000015 $ &  $ 7.74850 \pm 0.00022 $ \\
  $T_0$ [d]\tablenotemark{a} & $7744.07160 \pm 0.00022 $ &  $ 7751.1546 \pm 0.0010 $ \\
  $a$/\rstar & $2.292_{-0.060}^{+0.053}$  &   $21.59_{-0.74}^{+0.71}$ \\
  \rplanet /\rstar  & $0.02037 \pm 0.00046 $  & $0.094_{-0.037}^{+0.061} $ \\
  $i$ [deg] & $ 86.3_{-3.6}^{+2.7} (> 82.6)$ & $87.2_{-2.0}^{+1.6} $ \\
  \rplanet\ [\rearth] & $ 1.51 \pm 0.05 $ &  $7.0 ^{+4.6}_{-2.8}$ \\
\tableline                                 
$K$ [\ms ]\tablenotemark{b} & $6.25 \pm 0.48$  & $ < 3 $\tablenotemark{c} \\
$e$\tablenotemark{d} & 0 & 0 \\
$\omega$ [deg]\tablenotemark{d} & 90 & 90 \\
$\mathcal{M}_0$ [deg]\tablenotemark{b,e} & $ 182.2 \pm 0.6 $ & $ 238.5 \pm 0.1$ \\
\mplanet\ [\mearth]\tablenotemark{b} & $5.08 \pm 0.41$ & $< 7.4 $\tablenotemark{c} \\  
$\rho$ [\rhoearth ] & $1.48 \pm 0.20$ & \\
$\rho$ [${\rm g\, cm^{-3}}$] & $8.2 \pm 1.1$ & \\
\tableline
\end{tabular}
\tablenotetext{a}{Expressed as BJD$_{\rm TDB}$-2450000.0 d}
\tablenotetext{b}{Weighted average of the three methods}
\tablenotetext{c}{84.135$^{\rm th}$ percentile}
\tablenotetext{d}{Fixed}
\tablenotetext{e}{Mean anomaly at the reference time T$_{\text{ref}} = 7779.53438245$, \ie , the average of {\it K2} and HARPS-N epochs }
\end{table}

The confidence intervals of the posteriors of the fitted parameters are reported in Table~\ref{table:K2_parameter_planets}. The posterior distribution of the inclination of planet b is peaked at 90 degrees, hence we reported the 84.135$^{\rm th}$ percentile of the distribution from the peak as the lower limit on the inclination of the inner planet. The inclinations of the two planets are consistent with the two orbits being co-planar, although their posterior distributions peak at different values. The planet radii have been obtained using the stellar parameters in Section~\ref{sec:stellar_parameters}.

\section{Secondary Eclipse and Phase Curve of \planetb}\label{sec:secondary_analysis}
 We modeled the secondary eclipses and phase variations of \planetb\ using the {\tt spiderman} code \citep{Louden2017arXiv}. We used the primary transit parameters from Table~\ref{table:K2_parameter_planets} and the stellar \teff\ from Table~\ref{table:stellar_parameters}, with their uncertainties propagated. To account for the long exposure times we {\bf oversampled} the time series by a factor of 11 and then binned these values to get the final model points. The best-fitting parameters with their confidence intervals were obtained with an MCMC analysis using {\tt emcee} \citep[][described in the previous section]{ForemanMackey2013}. For each model we considered, we ran a 30 walker ensemble for 100,000 steps and checked for convergence, discarding the first 10,000 steps as burn in. 

Since the planet is so heavily irradiated, it is likely to possess an observable thermal flux in the visual, as well as a reflected light component. We first tested the plausibility of these two models independently, and then combined them.

For the reflection model we assumed that the planet reflects light uniformly as a Lambertian sphere, which translates to a geometric albedo in the {\it Kepler} bandpass. Since {\tt spiderman} models the phase curve and secondary eclipse simultaneously, the geometric albedo is the only additional model parameter over the primary transit model in the previous section.
We measured an occultation depth of $23\pm4$~ppm, meaning the secondary eclipse and phase signal are confidently detected at over 5$\sigma$ significance. The posterior for the geometric albedo has a mode and 68\% Highest Posterior Density (HPD)\footnote{Defined as the shortest possible interval enclosing 68\% of the posterior mass} interval of $0.30 \pm 0.06$. Such a high geometric albedo implies a relatively reflective atmosphere or surface, which would seem to be at odds with such a dense object orbiting so close to its star. The phase-folded data and the best-fitting reflection model are shown in Figure \ref{fig:secondary}

For the thermal model we used the simple physical model described in \citet{Kreidberg2016}, as implemented in {\tt spiderman}. 
The free parameters of this model are the planetary Bond albedo, and a day-to-night heat redistribution parameter, while the incident flux on the planet (required by the model) is obtained from the stellar and planetary parameters from the previous sections.
The mode and 68\% HPD interval for the Bond albedo is consistent with zero ($0.01^{+0.05}_{-0.01}$) with an upper limit of 0.37 (99.7$^{\rm th}$ percentile of the distribution). The redistribution factor is also consistent with zero, with a mode and 68\% HPD interval of $0.02^{+0.05}_{-0.02}$ and an upper limit of 0.23 (99.7$^{\rm th}$ percentile). This indicates a sharp day-night contrast, with a substellar surface temperature of 3000 K, or a surface averaged dayside value of $\sim$2400 K.

The maximum nightside temperature achievable in this model is 2100 K with maximum heat redistribution, which does not produce sufficient flux in the {\it Kepler} bandpass to be detected. 
However, this cannot rule out a nightside flux of different origin or a systematic underestimation of the total insolation of the planet. To test whether models with nightside flux might be preferred, we fitted the thermal model again, but added an extra free parameter to increase the temperature of the planet. This allows the freedom to fit a model with significant nightside flux, but the same occultation depth. We found no improvement to the fit, meaning there is no evidence for significant nightside flux in the data.

The reflected light and thermal models both produce fits that are comparable, both by eye and in terms of the dispersion of the residuals. Statistically speaking the former is preferred by the Bayesian Information Criterion ($\Delta {\rm BIC} = 12$), the latter having one extra degree of freedom, although the BIC alone is not sufficient to prefer one model over the other \citep[see for example ][for a review of the problems connected with the BIC]{Raftery99bayesfactors}, and a more careful model selection, possibly with the inclusion of new data at a different wavelength, should be performed.
To assess what can be said about the relative contributions of thermal and reflected light in the face of this model degeneracy, we ran a final combined model fit where the planet had both components. Since there is no evidence of nightside flux, we fixed the redistribution parameter of the thermal model to zero, thus these results should be seen as an upper limit.
The results of this analysis are shown in Figure~\ref{fig:contribution}, and show the 1$\sigma$ range of mutually acceptable Bond and geometric albedos. Since the signal can be reproduced satisfactorily with both pure thermal and pure reflective models, there is naturally a near perfect degeneracy. However, future observations at longer wavelengths should distinguish more easily between thermal and reflected light. This could set an upper limit on the Bond albedo for the planet, which would in turn break the degeneracy in this dataset and allow a more stringent upper limit to the geometric albedo to be set. 

\begin{figure}\includegraphics[width=\linewidth]{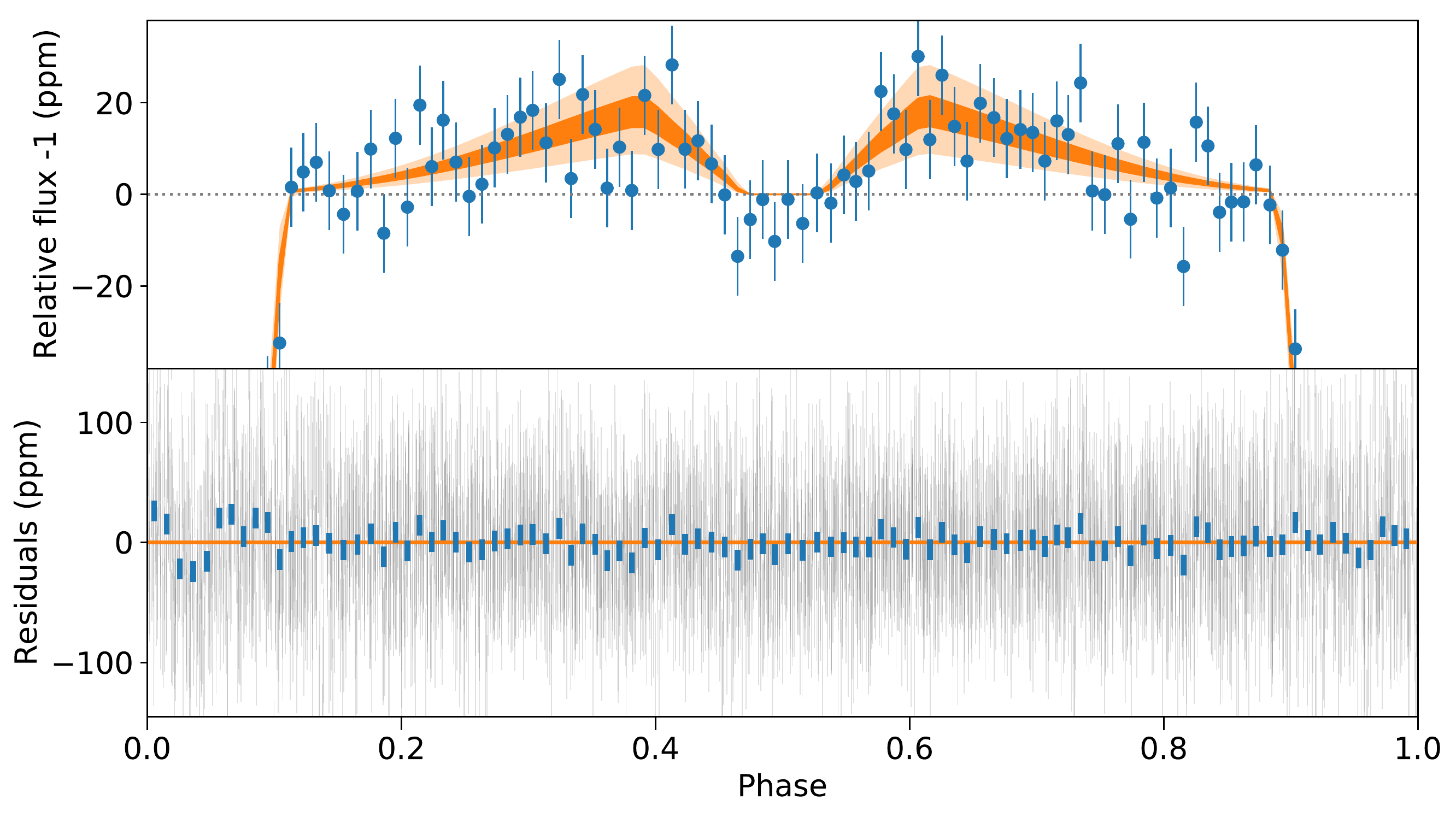}\caption{{\it Top:} The detrended data phase-folded on the period of planet b with the transits of planet c removed, the data have been binned by a factor of thirty for clarity. The size of the error bars is a model parameter, and is set by the maximum likelihood model. The 1 and 3 $\sigma$ credible intervals calculated from the posterior are overplotted in dark and light orange respectively. {\it Bottom:} The residuals to the best fitting model, the binned data are plotted as thick blue lines and the unbinned data is plotted as thin grey lines. All model fits were performed on the unbinned data.}\label{fig:secondary}\end{figure}

\begin{figure}\includegraphics[width=\linewidth]{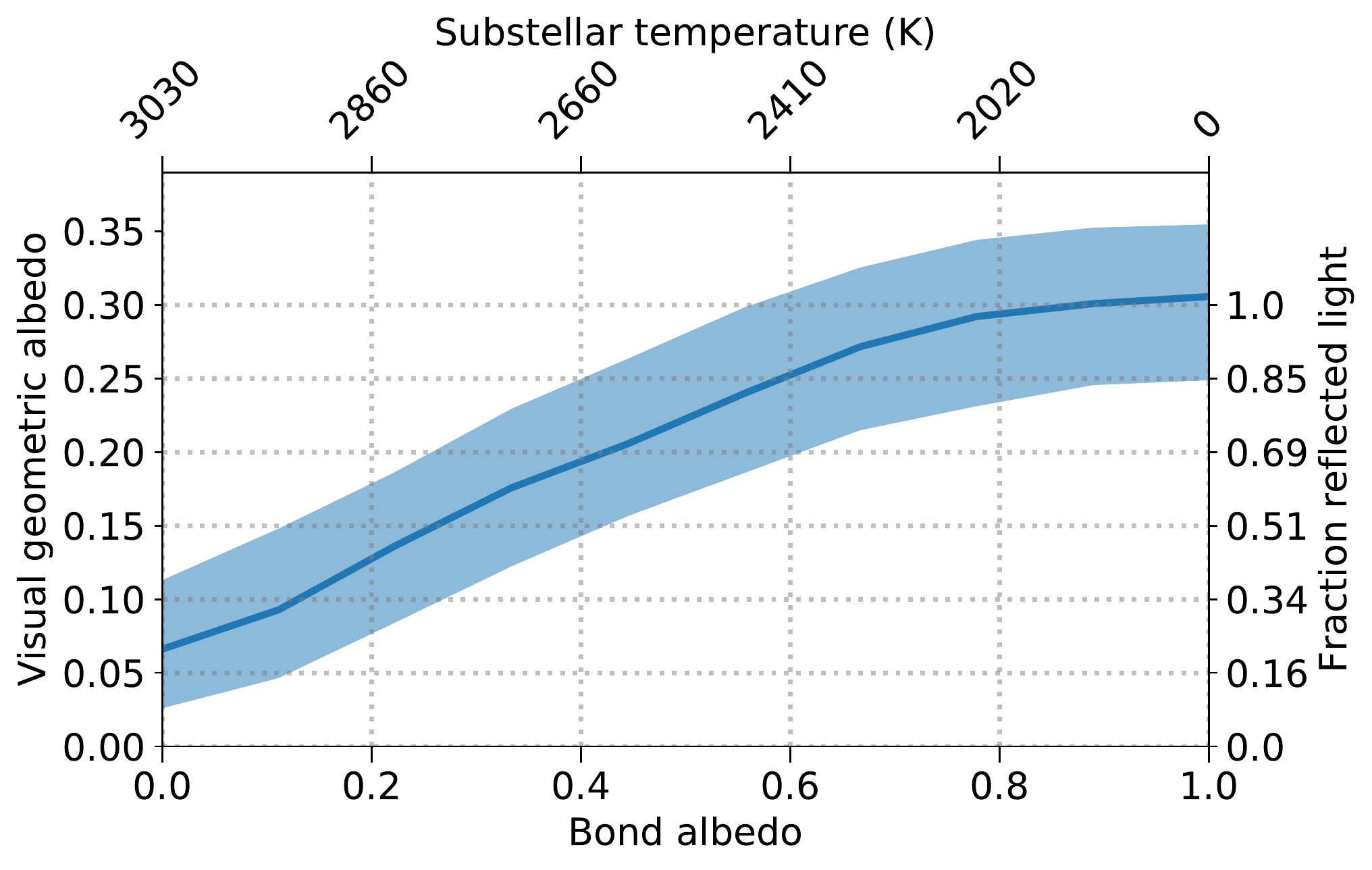}\caption{The best fitting visual geometric albedo as a function of the Bond albedo. The shaded area is the 68\% credible region for the geometric albedo, calculated using slices of the MCMC posterior. The corresponding substellar temperature for the planet is plotted on the top axis, and the fraction of the occultation depth from the reflected light alone is calculated using the best fitting Bond albedo for the corresponding visual albedo.}\label{fig:contribution}\end{figure}

\section{Radial Velocity Analysis}\label{sec:radial_velocity_analysis}

The twofold observational strategy we adopted to gather the RVs (see Section~\ref{sec:radial_velocities}), together with the high precision of the {\it K2} light curve and the availability of reliable spectroscopic  activity indices (thanks to the brightness of the star) allow us to model stellar activity with three different, complementary techniques, thus allowing an accurate determination of the planetary mass.

The analyses were performed assuming circular orbits for both planets. The circularization time scale for planet b is very short, given its short orbital period, and eccentricity excitation due to dynamical interactions with the outer planet are unlikely due to their separation in period, in addition to the fact that the architectures of USP systems seem dynamically cold \citep{Dai2017arXiv}. The adopted parameters are listed in Table~\ref{table:K2_parameter_planets} and correspond to the weighted average of the three techniques. In the following, confidence intervals are calculated by taking the 15.865$^{\rm th}$ and the 84.135$^{\rm th}$ percentiles of the posterior distributions, while upper limits are expressed as the 84.135$^{\rm th}$ percentile of the posterior.

\subsection{Nightly RV offsets}\label{sec:nightly-rv-offsets}
When the periodicity of the stellar activity is well separated from the orbital period of the planet, as in our case ($P_{\rm act} / P_{orb} \simeq 50 $), we can assume that the RV variation due to the activity, as well as the RV contribution from the outer planet, are constant within an orbital period of the USP planet. Since the nightly visibility window of our target from La Palma was shorter than the orbital period of the planet, this approach simply transforms into applying a nightly offset to our RV dataset. For this analysis we considered only those nights when at least two RVs were collected, for a total of 40 RVs across 15 nights. On the night of September 14th 2017, 9 consecutive RVs were gathered across 5.1 hours (0.21 days), almost covering a full orbital period.

We performed the analysis using the {\tt PyORBIT} code. Global optimization of the parameters was performed using the differential evolution code {\tt pyDE}\footnote{Available at \url{https://github.com/hpparvi/PyDE}}; the output was then fed to {\tt emcee} for a Bayesian estimation of the parameters and their errors. We used uninformative priors for all parameters except for the period of planet b, where we assumed a Gaussian prior with center and standard deviation set to the value and uncertainty obtained from the {\it K2} light curve (Section~\ref{sec:photometric_analysis}). The central time of transit was provided as input data. We used 80 walkers (4 times the number of free parameters) running for 50000 steps, of which the first 20000 were discarded as burn-in phase (although the Gelman-Rubin criterion for convergence was already met after a few thousand steps). After applying a thinning factor of 100 we were left with 24000 independent samplings for each parameter. We obtained an RV semi-amplitude of $K = 6.10 \pm 0.47$~\ms , corresponding to a planetary mass of \mplanet~$=4.96 \pm 0.39$~\mearth\ after taking into account the uncertainty on orbital inclination and stellar mass. The phase-folded RVs with their residuals are shown in the first panel of Figure~\ref{fig:three_techniques}.

\subsection{GPs and {\it K2} light curve}\label{sec:gps-k2-light}
The next approach assumes that light curve variations and activity signals in the RVs can be described by a GP with the same kernel and common hyper-parameters except for the covariance amplitude $h$, which is specific for each dataset.
We performed the analysis using the {\tt PyORBIT} code with the same kernel choice as described in Section~\ref{sec:stellar_activity}. As shown by \cite{Grunblatt2015} the quasi-periodic kernel is the best choice to model photometric and RV variations while preserving a physical interpretation of the hyper-parameters. 
We modeled the {\it K2} light curve and the RVs simultaneously, to better understand correlations between the hyper-parameters and the orbital parameters. We then repeated the analysis without including the {\it K2} light curve but using  the values obtained in Section~\ref{sec:stellar_activity} as priors on the hyper-parameters, with error bars enlarged by a factor of three to take into account a possible change in behavior of stellar activity during the time span between photometric and RV data. Differently from the previous approach, we included both planets in the model.

We ran the sampler for the same number of step as in Section~\ref{sec:nightly-rv-offsets}, using 68 walkers (four times the dimensionality of the model) for a total of 20400 independent samples when including the {\it K2} light curve, and 56 walkers for 16800 independent samples when imposing priors on the hyper-parameters. We followed the same criteria for convergence. The posteriors of the orbital parameters obtained in the two cases (\ie , with the {\it K2} light curve or imposing the priors) are nearly indistinguishable, \ie , we are not limited by the precise choice of the GP hyper-parameters.
For planet b we obtained an RV semi-amplitude of $K = 6.34 \pm 0.49 $ \ms , corresponding to a planetary mass of \mplanet $= 5.15 \pm 0.42 $ \mearth , while planet c was undetected, with a posterior distribution of $K_c$ peaked at zero and an upper limit of $K_c < 3.8 $ \ms . When including the {\it K2} data we obtained the same hyper-parameters as in Section~\ref{sec:stellar_activity} and a covariance amplitude $h_{\rm RV} = 11.4 \pm 2.5$~\ms , confirming the high level of activity of the star. The GP regression and the Keplerian contributions to the RVs are shown in the upper panel of Figure~\ref{fig:GP_comparison}. The phase-folded RVs with their residuals for planet b are shown in the second  panel of Figure~\ref{fig:three_techniques}. 

\subsection{GPs  and activity indices}\label{sec:gps-activity-indices}

In a third approach, we performed a combined analysis of RVs and activity indices using the GP framework introduced in \citet[][hereafter R15]{Rajpaul2015} and \citet{Rajpaul2016}. This framework was designed specifically to model RVs jointly with activity diagnostics even when simultaneous photometry is not available. It models both activity indices and activity-induced RV variations as a physically-motivated manifestation of a single underlying GP and its derivative.

We used R15's framework to derive a constraint on the activity component of the RVs, and joint constraints on the masses of planets b and c, independently of the approach based on the {\it K2} photometry. For this analysis, we modelled the \SHK, BIS and RV measurements simultaneously. Given strong observed linear correlations between \SHK\, CCF contrast and FWHM (Pearson correlation coefficients $\rho\sim0.8$, see Figure~\ref{fig:GLS_correlations}), modelling the latter two time series in addition to \SHK\ would have been redundant, as they would not have provided independent constraints on activity-induced RV variations.

We used a GP with quasi-periodic covariance kernel, as presented in R15, to model stellar activity, while we considered as a GP mean function either zero, one, or two non-interacting zero-eccentricity Keplerian signals (no planets, planet b only, and planets b and c) in the RVs only. We placed non-informative priors on all parameters related to the activity components of the GP framework (see R15); the priors we placed on the Keplerian orbital elements were the same as those in the preceding analyses.

We performed all parameter and model inference using the {\tt MultiNest} nested-sampling algorithm, with $2000$~live points and a sampling efficiency of $0.3$.

For planet b we obtained a RV semi-amplitude of $K_b = 6.31 \pm 0.49 $~\ms , corresponding to a planetary mass of $M_b= 5.14 \pm 0.42 $~\mearth\ (third panel of Figure~\ref{fig:three_techniques}), while planet c was again undetected but with a lower value on the upper limit for its RV semi-amplitude,  $K_c < 1.9 $~\ms .

We obtained GP hyper-parameters of $P_{\rm GP}=12.8\pm0.5$~d (overall period for the activity signal), $\lambda_{\rm p}=1.1^{+0.2}_{-0.1}$ (inverse harmonic complexity, with this inferred value suggesting an activity signal with harmonic content only moderately higher than a sinusoid), and $16^{+7}_{-5}$~d (activity signal evolution time scale). For the other hyperparameters we obtained $V_r = 66_{-13}^{+17}$~\ms\ and $V_c=-7.8_{-6.5}^{+5.3}$~\ms\ for the RVs ,  $L_c = 0.117_{-0.026}^{+0.033}$ for the \SHK\ index, $B_r = 30_{-8}^{+11}$~\ms\ and $B_c = -46_{-13}^{+10}$~\ms\ for the BIS. The best fit model is represented in the bottom panel of Figure~\ref{fig:GP_comparison}. Note that these parameters should not be compared directly with those reported in Sections~\ref{sec:stellar_activity} and \ref{sec:gps-k2-light}, as they are inferred based on fitting a combination of a quasi-periodic GP and its derivative to the RVs and multiple activity indices, while in the previous approach only the GP (without its derivative) is considered.

\begin{figure}
\includegraphics[width=\linewidth]{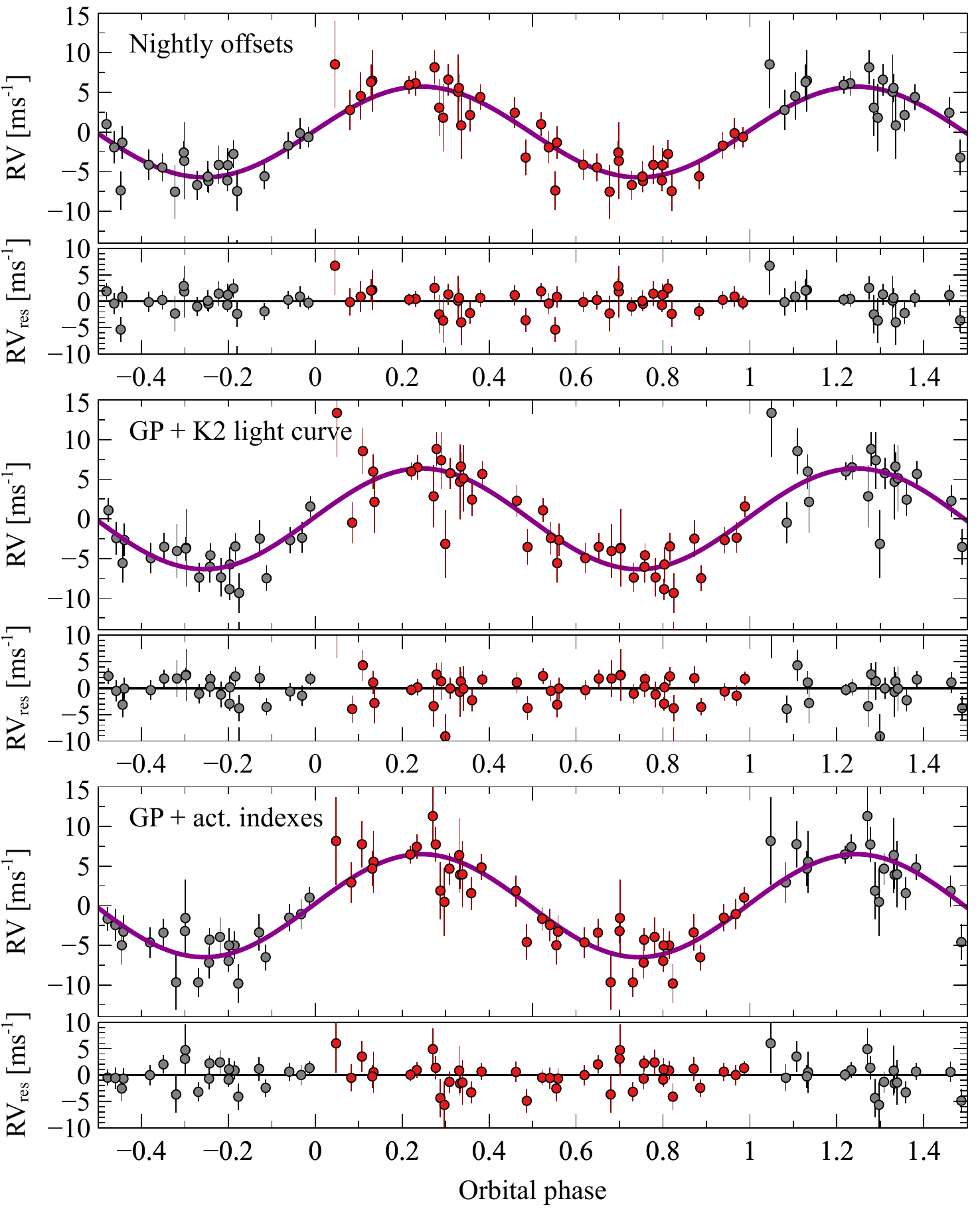}
\caption{Phase-folded RV fit with residuals for the three methods used in the analysis. For this plot we used the maximum {\it a posteriori} (MAP) parameter estimates. }
\label{fig:three_techniques}
\end{figure}

\begin{figure}
\includegraphics[width=\linewidth]{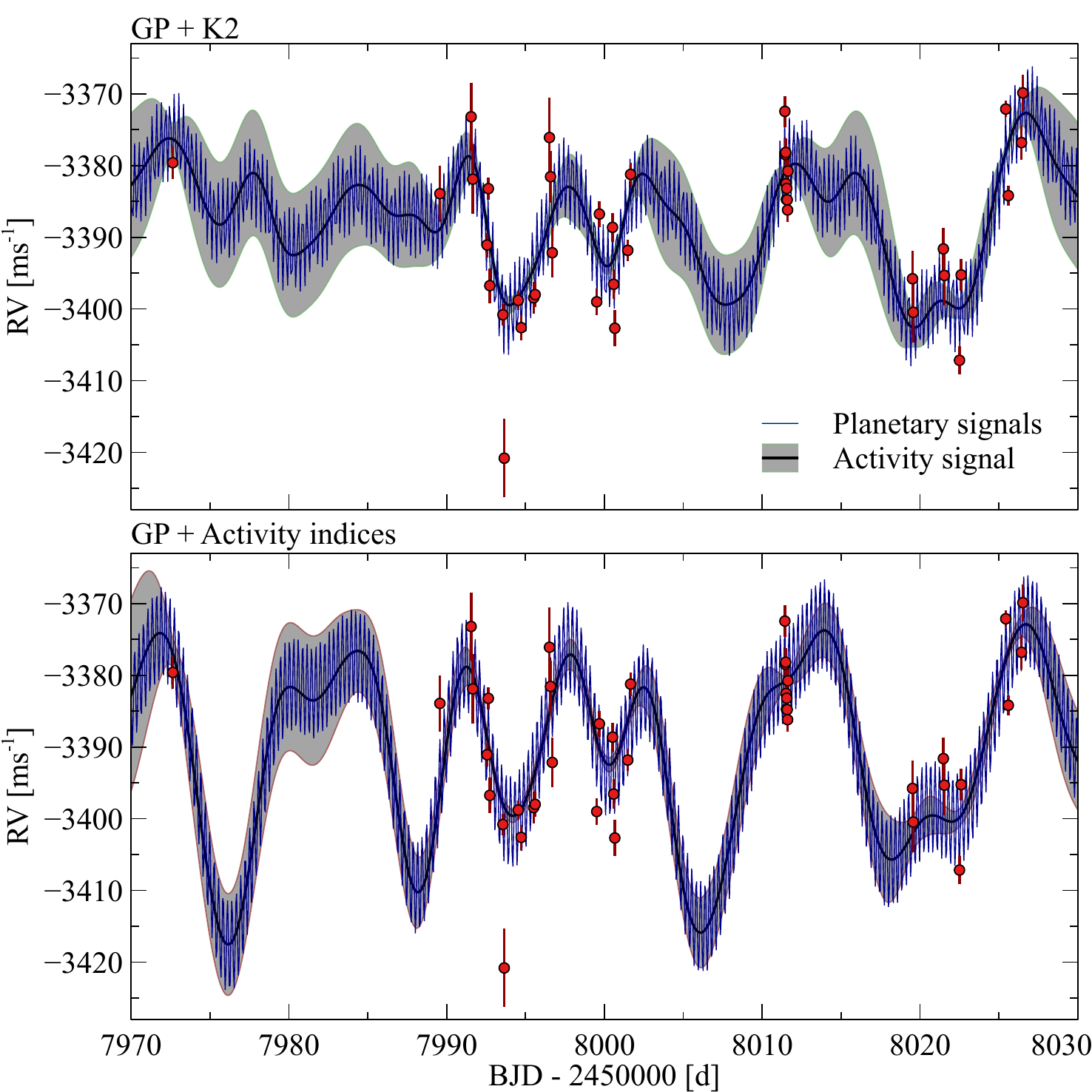}
\caption{Comparison of the combined stellar activity and planetary models obtained when using GP constrained by the {\it K2} light curve (upper panel) or the activity indices (lower panel). The blue curve represents the Keplerian contribute (using the MAP parameters), the black curve with the gray shaded area represents the GP regression with its 1$\sigma$ confidence interval.}
\label{fig:GP_comparison}
\end{figure}

\subsection{Effects of time integration}\label{sec:effects-time-integr}
For USP planets the variation of the RV curve during the time of one integration may become relevant. In our case, the exposure time of 1800 seconds (chosen to reach a good precision in RV) covers 8\% of the RV curve of the inner planet.  While this problem is not new in the exoplanet literature (\eg , in the analysis of the Rossiter-McLaughlin effect, \citealt{Covino2013}), it has never been addressed when dealing with RV fits for planet mass measurement. 
We proceeded as follows to estimate the systematic error in the semi-amplitude of planet b due to integration time: we computed a theoretical RV curve given the orbital parameters of the planet using a sampling of 180 seconds, then we binned this curve over ten points (corresponding to our integration time) and we measured the semi-amplitude $K_{\rm obs}$ of the resulting curve. By varying the input $K_{\rm true}$ we found that the $\Delta K = K_{\rm true}-K_{\rm obs}$, \ie , the correction to be applied to the observed semi-amplitude to recover the true value, is a linear function of $K_{\rm obs}$ with slope $9.07 \times 10^{-3}$ and null intercept. This suggests that the values of the semi-amplitude obtained by our fits are systematically underestimated by $\simeq  0.05$ \ms , \ie , well below the precision to which we can determine $K$. 
In a conceptually similar case involving a white dwarf orbiting a brown dwarf in a 91-minutes orbit, \cite{Rappaport2017} computed analytically the correction factor to be applied to an RV at a given epoch to take into account the finite exposure time (Equation~2 of their paper). By applying their equation, we obtain that the measured semi-amplitude is underestimated by a factor of $0.99$ ($\sim 0.06$~\ms ) with respect to the true value, in agreement with our previous estimate.

\subsection{The mass of planet c}\label{sec:mass-planet-c}
A commonly-encountered concern regarding GPs is that they may be flexible enough to wrongly `absorb' a planetary signal as stellar activity, resulting in a non-detection as in our case. Our GP-based methods are able to disentangle stellar signals from planetary ones even in cases where their periods are identical \citep[see e.g.][]{Mortier2016}, given that the latter would not in general have coherent phase and constant shape and amplitude over multiple stellar rotation periods. The GP component of the model is associated with a much higher complexity penalty than any Keplerian components, so the latter would be preferred regardless of the time span covered by the observations. In our case, the orbital period of planet c is close (but not identical) to the first harmonic of the rotational period of the star, so the previous considerations should remain valid. Nevertheless, we verified that with our tools we were always able to correctly retrieve injected RV signals with period and phase corresponding to planet c for several values of the semi-amplitude in the range between 1 and 20 \ms . This test also confirmed that our detection limit is not biased by the sampling of the observations.

In the previous section we carried out the RV analyses with a 2-planet model, motivated by the fact that we identified two planets in the {\it K2} light curve, and we confirmed that the semi-amplitude of planet c is consistent with zero using two complementary approaches to model stellar activity. The choice of the model can, however, strongly affect the outcome of the analysis \cite[see][for a recent example]{2017arXiv170706192R}; for example, the inferred parameters for planet b might be biased by the presence of a spurious second Keplerian term in the model, if indeed there is no detectable RV signal for planet c. We repeated the analyses by including only planet b in the model, and obtained posterior distributions for the orbital parameters and the GP hyper-parameters compatible with those of the 2-planet model, well within the 1$\sigma$ error bars. For the GP $+$ activity indices we also computed log model likelihoods (evidences) of $\ln \mathcal{Z}_0=-4.2\pm0.1$, $\ln\mathcal{Z}_1=33.3\pm0.1$ and $\ln\mathcal{Z}_2=33.8\pm0.1$ for the 0-, 1- and 2-planet models, respectively. On this basis we concluded that the model corresponding to an RV detection of planet b was favoured decisively over a zero-planet model, with a Bayes factor of $\mathcal{Z}_1/\mathcal{Z}_0\gtrsim10^{16}$. The Bayes factor $\mathcal{Z}_2/\mathcal{Z}_1\sim1.5$, however, indicated that there was no evidence to favour the more complex 2-planet model over the simpler 1-planet model.\footnote{We also considered non-circular orbits for planet c, but again this led to a non-detection. Moreover, the posterior distribution for the eccentricity was compatible with zero, with the simpler circular model being favoured with a Bayes factor $>10$.}

From our data, then, we are not able to recover the RV semi-amplitude of the outer planet. Our two GP-based modelling approaches yield different upper limits on the semi-amplitude of planet c, $K^{\rm GP+{\it K2}}_c < 3.9$ \ms\ versus  $K^{\rm GP+act}_c < 1.9$ \ms , possibly due in part to the different stellar rotational periods inferred by the two approaches, $P_{\rm rot}^{\rm GP+{\it K2}} = 13.9 \pm 0.2$~d versus $P_{\rm rot}^{\rm GP+act} = 12.8 \pm 0.5$~d, with the former being closer to twice the orbital period of the outer planet. It should be noted that the former rotational period is mainly driven by the {\it K2} photometry, which is more sensitive to the presence of starspots, while the latter is influenced by the \SHK\ index which probes the stellar chromosphere and is thus more sensitive to  the suppression of granular blueshift in magnetized regions of the star, as noted by \cite{Haywood2016}. The apparent discrepancy between the two measurements is likely due to the fact that we are sensing different physical effects.
From the posterior distribution of the semi-amplitude obtained in Section~\ref{sec:gps-activity-indices}, we can safely assume an upper limit for the RV signal induced by planet c of $K_{\rm c}=3 $ \ms\ (average of the upper limits obtained with two techniques), which translates into an upper limit on the mass of $\simeq 7.4$~\mearth .

\section{Discussion and conclusions}\label{sec:disc-concl}

We presented the validation and high-precision RV follow-up of two transiting planets discovered in the {\it K2} light curve of the very active star \starname . The innermost planet has a period of 0.28 days and falls into the category of so-called ultra-short period (USP) planets. 

We applied three independent but complementary approaches in an attempt to minimize the effects of our assumptions when modelling the stellar activity signals. Namely, we used the nightly offsets to remove all the signals with time scale larger than the period of the inner planet; a GP approach where the values of the  hyper-parameters are mostly driven by (non-simultaneous) high precision photometry; and a GP approach where the simultaneous activity indices are modelled with the same underlying model for the stellar activity in the RVs, without relying on photometry.
Figure~\ref{fig:GLS_residuals} shows the lack of correlation of activity indices with the RVs after removing only the activity, \ie , there is no correlation between the planetary signals and the activity indices.
Notably, the three complementary methods all yielded the same conclusions, resulting in a mass measurement for the innermost planet that is not only precise but also robust. The nightly offset approach resulted in a slightly smaller semi-amplitude of planet b ($K_b = 6.1 \pm 0.5$~\ms ) with respect to the GP approaches ($K_b = 6.3 \pm 0.5$~\ms ), 
well within the error bars. \planetb\ is thus confirmed at over 12$\sigma$ confidence.
We measured a radius of $R_b = 1.51 \pm 0.05 $~\rearth\ from {\it K2} light curve and a mass of $M_b = 5.1 \pm 0.4$~\mearth\ from HARPS-N spectra, resulting in a density of $\rho _b = 1.48 \pm 0.20 $~\rhoearth\ $ = 8.2 \pm 1.1$~${\rm g\, cm^{-3}}$. 

\planetb\ joins the small sample of USP planets with precisely known masses and radii, shown in Figure~\ref{fig:mass_radius_diagram}: 55 Cnc e \citep{Mcarthur2004, Winn2011, Demory2011, Nelson2014, Demory2016b}, CoRoT-7b \citep{Leger2009, Queloz2009, Haywood2014}, WASP-47e \citep{Becker2015, Dai2015, Sinukoff2017b,Vanderburg2017}, Kepler-78b \citep{SanchisOjeda2013, Pepe2013, Howard2013,Hatzes2014, Grunblatt2015}, Kepler-10b \citep{Batalha2011,Dumusque2014,Weiss2016,2017arXiv170706192R}, K2-131b \citep{Dai2017arXiv}, HD3167b \citep[][respectively labeled as  C17 and G17 in the plot]{Vanderburg2016c, Christiansen2017,Gandolfi2017}, K2-106b \citep[][S17 and G17 respectively]{Sinukoff2017,Guenther2017}. Notably, the last two planets have two independent density measurements which are not consistent with each other, resulting in a disagreement in the interpretation of the internal composition. The density of \planetb\ is consistent with a rocky terrestrial compositions, \ie ,  mainly silicates and iron, most probably with a large iron core between 30\% and 50\% of the total mass. From our density estimate we can exclude the presence of a thick envelope of volatiles or H/He on the surface of the planet.

\begin{figure*}
\includegraphics[width=\linewidth]{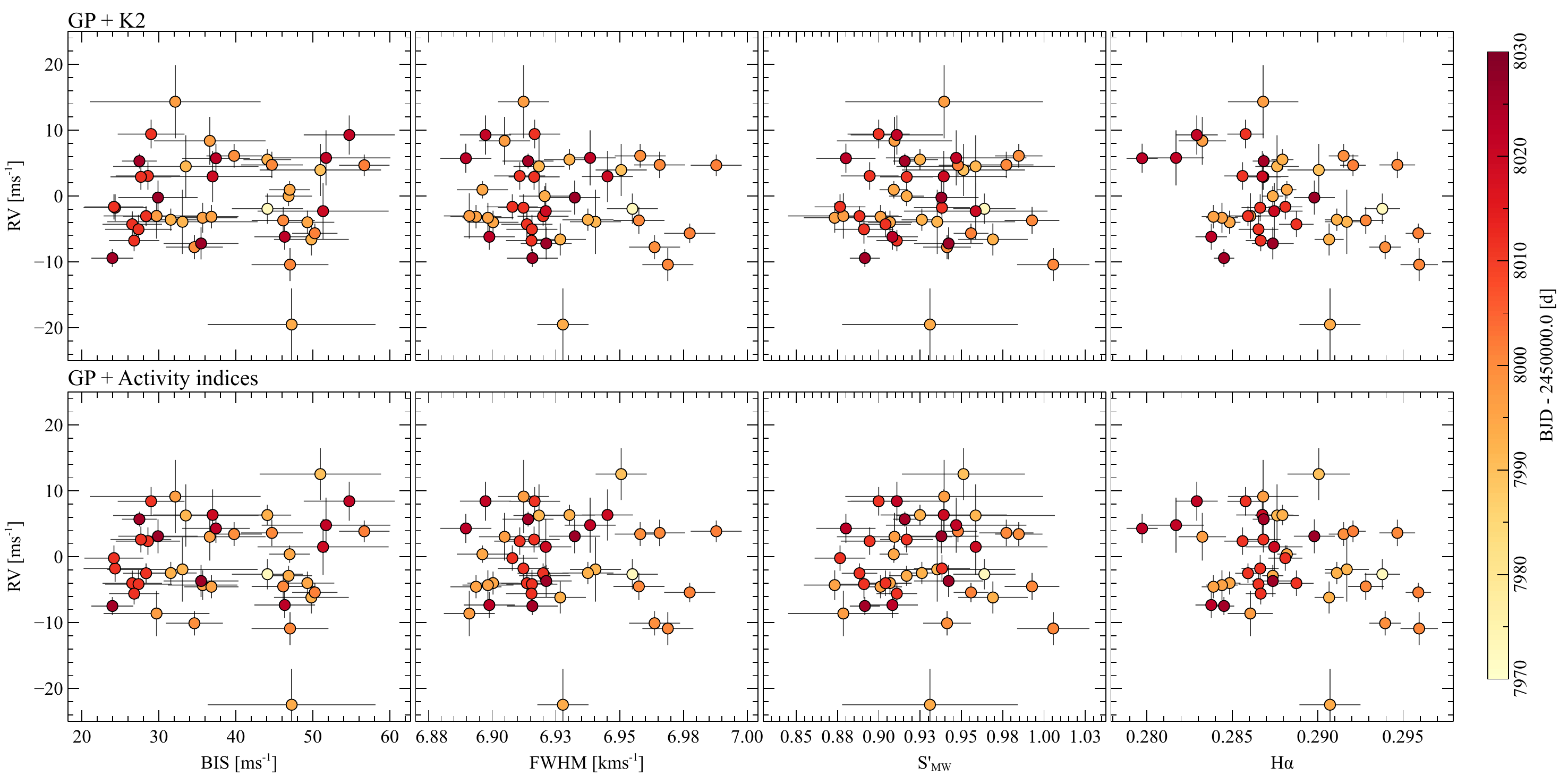}
\caption{Activity indices versus radial velocity, after removing the contribution of activity from the latter. We used the same limits as Figure~\ref{fig:GLS_correlations}.}
\label{fig:GLS_residuals}
\end{figure*}

\begin{figure*}
\includegraphics[width=\linewidth]{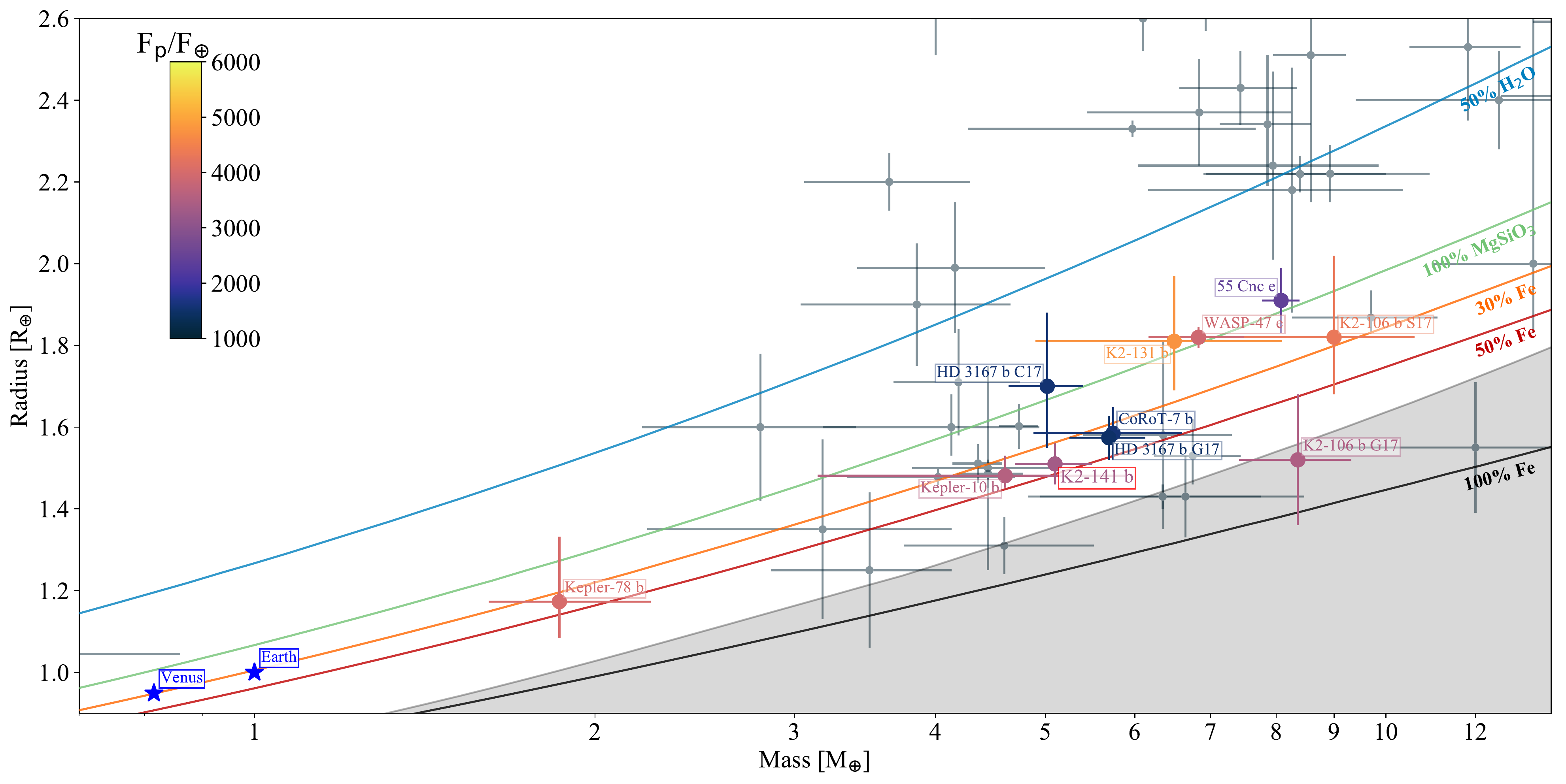}
\caption{Mass-radius diagram for the known USP planets, color-coded according to their incident flux. Grey points represent planets with period longer than one day and mass measurement more precise than 30\%.}
\label{fig:mass_radius_diagram}
\end{figure*}

We detected and analyzed the secondary eclipse and phase curve variation of planet b. The data is compatible with either a thermal emission of 3000~K from the day-side and an upper limit of 0.37 (99.7$^{\rm th}$ percentile) on the Bond albedo, or a planet with geometric albedo of $0.30 \pm 0.6$. The {\it Kepler} bandpass does not allow us to distinguish between the two models, with the truth characteristics of the planet probably lying between the two models. Infrared observations with the {\it Hubble Space Telescope} and the forthcoming {\it James Webb Space Telescope} will be able to refine the Bond albedo and thus constrain the geometric albedo of the planet.

The second planet has a period of around 7.75 day and since its transits are grazing, its radius cannot be measured precisely ($R_p =  7.0 ^{+4.6}_{-2.8}$ \rearth). The mass of \starname c is also not measured precisely because the planet's orbital period is close to the first harmonic of the rotational period of the star and/or because its RV signal may simply be too small to detect. From our dataset we were only able to put an upper limit on the planet's mass of $\simeq 8$~\mearth\ (84.135$^{\rm th}$ percentile of the distribution). Due to the weak constraints on the mass and radius of planet c, we are not able to shed much light on its likely composition, but our mass limit suggests that the planet is more likely a mini-Neptune or a Neptune-like planet with a thick envelope than a rocky planet or a HJ. The discovery of a second planet in a grazing configuration, initially missed by automatic pipelines, corroborates the previously observed trend that USP planets are often found in multi-planet systems \citep{SanchisOjeda2014}.

\begin{acknowledgements}

The HARPS-N project was funded by the Prodex Program of the Swiss Space Office (SSO), the Harvard- University Origin of Life Initiative (HUOLI), the Scottish Universities Physics Alliance (SUPA), the University of Geneva, the Smithsonian Astrophysical Observatory (SAO), and the Italian National Astrophysical Institute (INAF), University of St. Andrews, Queen's University Belfast and University of Edinburgh.

The research leading to these results received funding from the European Union Seventh Framework Programme (FP7/2007- 2013) under grant agreement number 313014 (ETAEARTH).

VMR acknowledges the Royal Astronomical Society for financial support.

MHK acknowledges Allan R. Schmitt for his continuous effort to develop LcTools.

This work was performed in part under contract with the California Institute of Technology/Jet Propulsion Laboratory funded by NASA through the Sagan Fellowship Program executed by the NASA Exoplanet Science Institute.

This work has been carried out in the frame of the National Centre for Competence in Research PlanetS supported by the Swiss National Science Foundation (SNSF). DE, CL, FB, DS, FP and SU acknowledge the financial support of the SNSF. DE acknowledges support from the European Research Council (ERC) under the European Union’s Horizon 2020 research and innovation programme (project {\sc Four Aces}; grant agreement No 724427).

PF acknowledges support by Funda\c{c}\~ao para a Ci\^encia e a Tecnologia (FCT) through Investigador FCT contract of reference IF/01037/2013/CP1191/CT0001, and POPH/FSE (EC) by FEDER funding through the program ``Programa Operacional de Factores de Competitividade - COMPETE''. PF further acknowledges support from FCT in the form of an exploratory project of reference IF/01037/2013/CP1191/CT0001. 

Parts of this work have been supported by NASA under grants No. NNX15AC90G and NNX17AB59G issued through the Exoplanets Research Program.

Based on observations made with the Italian Telescopio Nazionale Galileo (TNG) operated on the island of La Palma by the Fundación Galileo Galilei of the INAF (Istituto Nazionale di Astrofisica) at the Spanish Observatorio del Roque de los Muchachos of the Instituto de Astrofisica de Canarias.

This research has made use of the SIMBAD database, operated at CDS, Strasbourg, France

This publication makes use of data products from the Two Micron All Sky Survey, which is a joint project of the University of Massachusetts and the Infrared Processing and Analysis Center/California Institute of Technology, funded by the National Aeronautics and Space Administration and the National Science Foundation.

This research has made use of the Exoplanet Follow-up Observation Program website, which is operated by the California Institute of Technology, under contract with the National Aeronautics and Space Administration under the Exoplanet Exploration Program.  

This paper includes data collected by the Kepler mission. Funding for the Kepler mission is provided by the NASA Science Mission directorate.

Some of the observations in the paper made use of the NN-EXPLORE Exoplanet and Stellar Speckle Imager (NESSI). NESSI was funded by the NASA Exoplanet Exploration Program and the NASA Ames Research Center. NESSI was built at the Ames Research Center by Steve B. Howell, Nic Scott, Elliott P. Horch, and Emmett Quigley. 

\end{acknowledgements}

\software{LcTools \citep{Kipping2015}, DRS, CCFpams \citep{Malavolta2017b}, ARES (v2; \citealt{Sousa2015}), MOOG \citep{Sneden1973}, ATLAS9  \citep{Castelli2004}, SPC \citep{Buchhave2012, Buchhave2014}, isochrones \citep{Morton2015}, MultiNest \citep{Feroz2008,Feroz2009,Feroz2013}, MIST \citep{Dotter2016,Choi2016,Paxton2011}, PyORBIT (v5; \citealt{Malavolta2016}), george \citep{Ambikasaran2015}, vespa \citep{Morton2015b}, emcee \citep{ForemanMackey2013}, spiderman \citep{Louden2017arXiv}, pyDE \citep{Parviainen2016}}


\bibliographystyle{aasjournal} 
\bibliography{Bibliography.bib} 

\end{document}